\documentclass[
preprintnumbers,amsmath,amssymb,superscriptaddress]{revtex4}
\usepackage{graphicx}

\begin{document}

\title{The Neutron EDM Experiment}
\author{Philip Harris}
\affiliation
{Department of Physics and Astronomy, University of Sussex, Falmer, Brighton BN1 9QH, UK }


\begin{abstract}
The neutron EDM experiment has played an important part over many decades in shaping and constraining numerous models of CP violation.  
This review article discusses some of the techniques used to calculate EDMs under various theoretical scenarios, and highlights some of the implications of EDM limits upon such models.  A pedagogical introduction is given to the experimental techniques employed in the recently completed ILL experiment, including a brief discussion of the dominant systematic uncertainties.  A new and much more sensitive version of the experiment, which is currently  under development, is also outlined.
\end{abstract}
\maketitle

\section{Introduction}

\begin{figure}[ht]
  \begin{center}
    \resizebox*{0.6\textwidth}{!}{\includegraphics
    {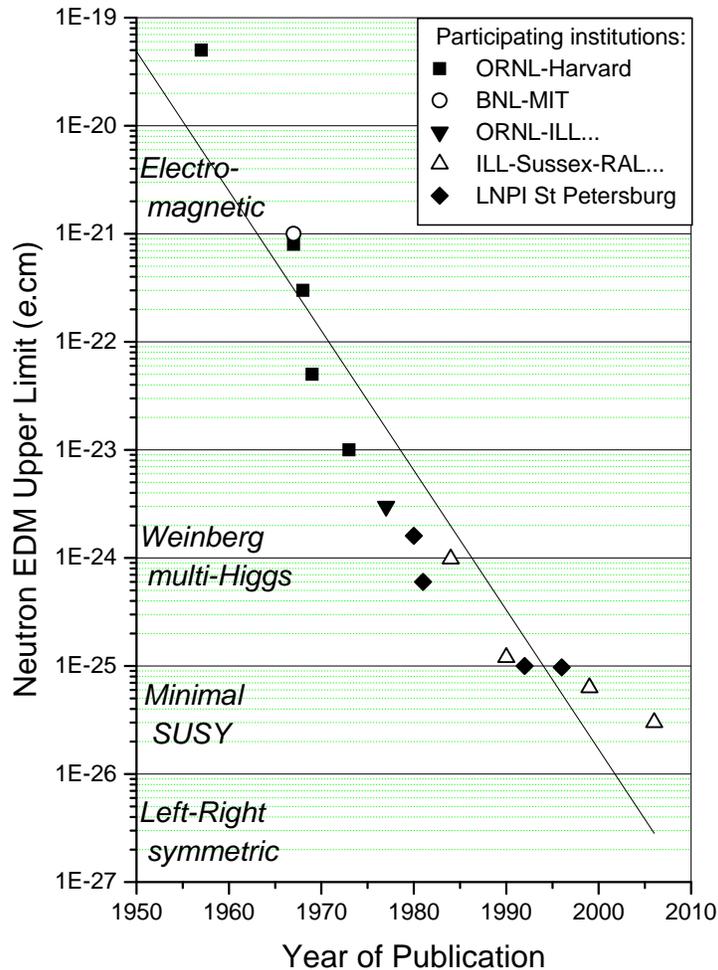}}
    \caption{Sensitivity of neutron EDM experiments over time.  On the left of the graph are some theoretical predictions of the magnitude of the neutron EDM. 
}
    \label{fig:edmlimit}
  \end{center}
\end{figure}

It is now more than half a century since Ramsey carried out his pioneering first search for a permanent electric dipole moment (EDM) of the neutron \cite{smith57}.  Ramsey's original motivation \cite{purcell50}, in contravention to the prejudices of the day, was to test the conservation of parity. The null result obtained was not believed to be important, and it lay unpublished until the aftermath of the 1956 discovery of parity violation in the radioactive decay of $^{60}$Co \cite{wu57}.  The implications for $T$ violation were then realised, at which point the search picked up -- and has continued ever since, achieving an impressive order of magnitude improvement in sensitivity every eight years or so \cite{pendlebury_hinds00}, as shown in Figure \ref{fig:edmlimit}. 
 To put the extraordinary degree of sensitivity of this experiment into a human perspective, if one imagines a neutron expanded to the size of the Earth, the current limit of $|d_n| < 2.9\times10^{-26}$ $e$ cm\cite{baker06, baker07} would correspond to a separation of the centres of charge of about 1.5 microns.  (Equivalently, there are 27 orders of magnitude between the size of the EDM limit and the size of a football -- the same scale factor as between the size of a football and the size of the visible universe.)  In energy terms, one can regard this experiment as having the ability to detect a level splitting of less than 10$^{-21}$ eV.  

It is easy to visualise why the existence of a permanent EDM $\vec{d}$ in a fundamental particle  violates both parity $P$ and time-reversal symmetry $T$.   The EDM itself is a vector proportional to the neutron's spin (i.e., either parallel or antiparallel to it -- there being no other direction defined for a spin-1/2 particle).  Therefore, when such a particle is placed within an electric field $\vec{E}$, a $P$-odd, $T$-odd term $-\vec{d}\cdot\vec{E}$  is introduced into the Hamiltonian: A parity reversal changes the sign of $\vec{E}$ but leaves the spin (and hence the EDM) unchanged, whereas the opposite is true under $T$ reversal.  In some other systems, this is perfectly acceptable; the ammonia molecule, for instance, exists in two degenerate mirror-image forms, and an EDM is therefore possible without parity violation.  The reality of the Pauli principle applied to nuclear structure, however, tells us that there is one, and only one, type of neutron, so the existence of a permanent EDM would be a direct manifestation of $P$- and $T$- violation.

It is worth noting that a fallacious argument relating to the charge structure of the neutrons is often employed. 
One can imagine the neutron to be a particle that is electrically neutral overall, but with positive and negative centres of charge slightly offset from one another along the spin axis.  It appears superficially that a parity reversal would swap the charge distribution while leaving the spin direction unchanged, and that parity would therefore be violated; however, since $\vec{E}$ changes sign under $P$, in this scenario the sign of $-\vec{d}\cdot\vec{E}$ would not change, and no parity violation would occur.  Instead, we must regard the EDM as being coupled to the spin itself, rather than being independent of it.

The neutron EDM is also deeply related to the question of the baryon asymmetry of the Universe \cite{trodden99}.  The Standard Model parameterization of CP violation is unable to accommodate the relatively high value of the baryon-to-photon ratio $\eta = 6\times10^{-10}$.  Ellis \cite{ellis81a} describes how, in a wide class of GUTs, the diagrams that generate $\eta$ also contribute to the renormalization of the CP-violating QCD parameter $\theta$ (see below), and hence to the neutron EDM; in a subsequent paper \cite{ellis81b}, Ellis describes the neutron EDM as a ``sensitive cosmological seismometer''.  For the GUTs in question, one obtains the order-of-magnitude inequality $d_n \stackrel{\scriptstyle <}{\scriptstyle\sim} 2.5\times 10^{-18} \eta e$cm, which suggests that in an almost model-independent way the neutron EDM should lie within about an order of magnitude of the existing experimental limit.

To date, no permanent EDM of any fundamental particle has yet been found, and the ongoing measurements continue in their long tradition of placing ever-tighter constraints upon $CP$-violating theories beyond the Standard Model.  

\subsection{Modelling the neutron EDM}

It is appropriate to ask what size of EDM one might expect, depending upon the nature of the $T$-violating forces that underlie it.  From dimensional arguments, one can write \cite{perkins}
\begin{center}
EDM  = charge ($e$) $\times$ a length ($l$) $\times$ a $T$-violation parameter $f$.
\end{center}
If the strong force were $P$- and $T$-violating, one might expect $lf$ to be a reasonable fraction of the diameter of the neutron, or perhaps of its Compton wavelength $\hbar/Mc$, giving an EDM of the order 10$^{-14}$ $e$ cm in size.  This is not the case, however, and we expect to have to bring in the weak interaction to accommodate parity violation.  A ``natural'' length scale that suggests itself is then $l = GM$, where $G = 10^{-5}/M^2$ is the weak-interaction coupling constant, thus giving $d_n \sim 10^{-19} fe$ cm.  Studies of neutral kaon decays \cite{christenson64} suggest that CP- (and hence T-) violating forces are a thousand times weaker than the weak force, thus implying that $f  \stackrel{\scriptstyle <}{\scriptstyle\sim} 10^{-3}$ and leading to an expected upper limit of $d_n \sim 10^{-22}\ e$ cm.  

Ellis \cite{ellis89} provides a good review of the techniques actually used to model EDMs.  The neutron EDM is often calculated on the basis of the usual vectorial addition of the contributions expected from the constituent quarks in the na\"ive quark model (NQM): $d_n = \frac{4}{3}d_d - \frac{1}{3}d_u$.  One can calculate contributions from the colour dipole moment in a similar manner \cite{pospelov01}.  In a later paper, going beyond the NQM, Ellis \cite{ellis96} points out that the spin of the neutron also has a substantial contribution from the strange quark, which can provide significant cancellation of the otherwise-dominant $d$-quark contribution.  A phenomenological approach is also often adopted.  One interesting example of this is given by He et al.\ \cite{he88}, who within the framework of left-right symmetric models relate $d_n$ to the measured value of $\epsilon'/\epsilon$, and who thereby predict $d_n \geq 1.9\times10^{-27}\ e$cm, which should be well within the range of the next generation of experiments.  

\subsection{Standard Model Predictions}

The simplest way in which EDMs can arise in the Standard Model \cite{dar00} is through W exchange.  Study of the contributing tree-level W-exchange diagrams reveals a simple but important property: Because there is no net flavour change, all first-order contributions to 
an EDM must cancel.  These processes of necessity involve sums of matrix elements $V_{ij}^*V_{ij}$ where $i, j$ are quark flavours.  These contributions are real, and therefore do not involve the $CP$-violating phase $\delta$.  A finite EDM from the Standard Model therefore requires second-order interactions: an additional gluon loop, or an interaction involving more than one quark, will lead to contributions at this level.

Donoghue et al.\ \cite{donoghue} give an estimate, based on simple dimensional considerations, of the expected neutron EDM arising from the second-order contributions within the Standard Model.  As a consequence of the required invariance of results under rephasing of quark fields, all $CP$-violating observables must include a factor of 
\[{\rm Im} \Delta^{(4)} = A^2\lambda^6\eta = c_1c_2c_3s_1^2s_2s_3\delta \sim 10^{-4},\]
where $A, \lambda, \eta, c_1$ etc.\ are from the usual Wolfenstein/CKM parameterisations.  In addition, the GIM mechanism causes the contributions of degenerate-mass quarks to cancel, so one expects contributions of the form $(m_j^2-m_k^2)/M_W^2$, dominated by the $m_t$ term.  In total, one expects
\[d_n \sim e\frac{G_F^2}{\pi^4}\frac{m_t^2}{M_W^2}{\rm Im~}\Delta^{(4)}\mu^3\sim 10^{-31}\ e {\rm cm},\]
where $\mu$ is a ``typical hadronic scale'' (0.3 GeV was used) included to make the dimensions correct and the factors of $\pi$ are anticipated from loop diagrams.  This estimate falls within the range of 10$^{-33}$-10$^{-30}$ $e$cm predicted from a range of more sophisticated calculations.  This is several orders of magnitude below the current limits, and is beyond the sensitivity of any current or foreseen technology.

From an experimentalist's point of view, the importance of this result is that any signal of an EDM is {\em background free}.  More conventional (high-energy) experiments exploring CP violation, such as those at the SLAC and KEK B factories \cite{babar04, belle04}, have to search for small departures from a relatively large anticipated signal.  In EDM searches, once systematics are overcome and a definite signal is found, it can immediately and definitely be recognised as a signal of physics beyond the Standard Model.  Furthermore, as most theories beyond the Standard Model do allow first-order CP violation they tend to predict EDMs that are close to the experimental bounds.

\subsection{The Strong $CP$ problem}

The QCD Lagrangian includes an arbitrary $P$- and $T$-violating phase $\theta$.  Furthermore, in the Standard Model this phase is shifted from its ``raw'' value: When the Higgs field acquires its vacuum expectation value, the Yukawa couplings with the fermions produce the quark mass matrices.  The diagonalisation of these matrices produces the aforementioned shift in $\theta$, to yield
\[\bar{\theta} = \theta + {\rm arg}({\rm det~}{\bf m'}),\]
where ${\bf m'}$ is the original (nondiagonal) mass matrix.  

The $\bar{\theta}$ term occurs in a $\Delta S = 0$ operator, and thus is not responsible for the $CP$ violation observed in $K$ decays.  However, it does generate a neutron EDM of order \cite{donoghue}
\[d_n \simeq \bar{\theta}\frac{m_um_dm_s}{m_um_d+m_um_s+m_dm_s}\frac{e\mu_n}{\Delta M} \sim \bar{\theta}\times 10^{-15}\: e {\rm cm},\] 
where $\mu_n$ is the nuclear magneton and the $m_i$ are the quark masses.  (Note that $\bar{\theta}\rightarrow 0$ naturally if any of the quarks has zero mass; however, this is disfavoured phenomenologically.)  Existing limits on the neutron (and also on $^{199}$Hg \cite{romalis01}) EDM therefore constrain this otherwise apparently arbitrary phase to be less than 10$^{-10}$ radians.  This anomaly is known as the Strong $CP$ problem.

The Strong $CP$ problem is not satisfactorily resolved within the Standard Model.  Perhaps the best candidate mechanism for the suppression of $\bar{\theta}$ is that proposed by Peccei and Quinn, who suggest the imposition of an additional symmetry that constrains $\bar{\theta}$ to be zero \cite{peccei77}.  This additional symmetry, however, requires in its turn the existence of an as-yet-unobserved particle, the axion, which has been sought without success for many years \cite{collar03, tada01}.

\subsection{Contributions from SUSY}

\begin{figure}[ht]
  \begin{center}
    \resizebox*{0.4\textwidth}{!}{\includegraphics
    {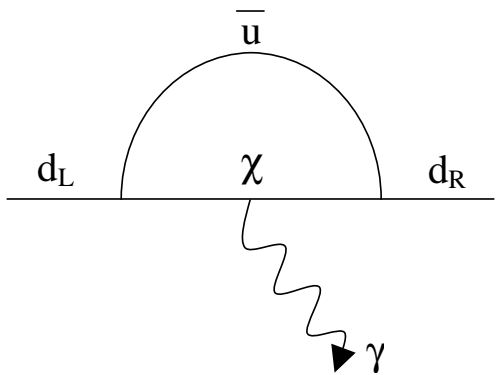}}
    \caption{A supersymmetric contribution to the neutron EDM
}
    \label{fig:susy_diag}
  \end{center}
\end{figure}

Supersymmetry postulates the existence of a fermionic (bosonic) `superpartner' for every boson (fermion) in the Standard Model.  Phases in the gluino mass and in the squark mass matrix then provide new contributions to the neutron EDM, such as that shown in Figure \ref{fig:susy_diag}. 

In order to evaluate these contributions, one has to know the masses of the squarks and the couplings at the vertices; these depend upon the details of the supersymmetry breaking for the model under consideration.   If the parameters are chosen appropriately, electroweak symmetry is broken naturally with the right $W$ mass.  However, this typically leads to quark EDMs that are of order
\[d_q\sim ({\rm loop~factor,~typ.~1}/\pi)\times\frac{m_q}{\Lambda^2}\sin\Phi_{CP},\]
where $m_q$ is the quark mass, $\Lambda$ is the scale of SUSY breaking ($\stackrel{\scriptstyle >}{\scriptstyle\sim}$ 200 GeV) and $\Phi_{CP}$ is a $CP$-violating phase, naturally of order 1; this in turn yields values for the neutron EDM of about $d_n\sim10^{-23}$-$10^{-24}\ e$cm, which is a factor of 100 to 1000 times too large. (Colour dipole moments, where the photon in Figure \ref{fig:susy_diag} is replaced by a gluon, are also ``naturally'' predicted to be large; these contribute substantially to the calculated $^{199}$Hg EDM).  The EDMs can be suppressed, by suppressing the phase $\Phi_{CP}$ or by making the symmetry-breaking scale $\Lambda$ (unattractively) large, 
but the fine-tuning of parameters in this way is rather undesirable; and, furthermore, suppression of the $CP$-violating phase would probably require a different source of $CP$ violation for baryogenesis.  Cancellations may also be invoked, but it then becomes difficult to incorporate the existing EDM limits from all of the systems currently under study -- neutron, electron and $^{199}$Hg.  A review of this situation is given by Abel et al.\ \cite{abel01}.  Detailed predictions of the EDM of $^{199}$Hg arising from the Minimal Supersymmetric Standard Model (MSSM), and conversely of the constraints on MSSM from $^{199}$Hg EDM measurements, are given by Falk et al.\ \cite{falk99}.

\subsection{Implications of non-zero EDM measurements}

EDMs are being sought in various systems: the free neutron, the mercury atom \cite{romalis01}, and the electron (via the thallium atom \cite{regan02} and, more recently, the YbF \cite{hudson02} and PbO molecules \cite{kozlov02}), in addition to a recent proposal to study deuterium \cite{lebedev04}.  The fundamental mechanisms underlying sources of EDMs are different in each system, and the measurement of a finite value within one of these systems would therefore have distinctive implications \cite{abel06}: For example, if the edms are driven by the QCD  $\theta$ angle, one 
would expect similar contributions to all strongly coupled systems, with 
the consequent approximate pattern $d_n\sim d_{Hg}\sim d_D >> d_{e},$ and so on.  Thus, the different systems have different implications for physics models beyond the standard model. Measurements on multiple systems are also needed in order to rule out cancellations.

From the foregoing, it is clear that EDM limits provide fairly tight constraints upon supersymmetric models; the same is true of most other models beyond the standard model that attempt to incorporate $CP$ violation to a degree adequate to explain the observed baryon asymmetry of the Universe.  The ``accidental'' cancellation of first-order contributions in the Standard Model is not a general feature, and EDM limits (and EDM values, once measured) provide a powerful way to distinguish between models and, indeed, to eliminate many of them.  Ramsey \cite{ramsey95} and Barr \cite{barr93a} have provided useful reviews of the situation, and the book by Khriplovich and Lamoreaux \cite{khriplovich_lamoreaux} contains further general information on EDMs.  An up-to-date review of CP violation, including EDM constraints, is provided by Ibrahim and Nath \cite{ibrahim07}. 

\section{Experimental technique}

An EDM is detected via the tiny Stark splitting induced by an electric field, which is applied alternately parallel and antiparallel to a small applied magnetic field.  A change in the neutron's resonant (Larmor precession) frequency, proportional to the electric field, is indicative of the presence of an EDM.  Specifically, in the presence of parallel (antiparallel) magnetic and electric fields, the resonant frequency $\nu$ is given by
\[h\nu = 2 \left(\mu_BB \pm d_nE\right),\]
and so the frequency difference between the parallel and antiparallel cases (assuming no change in $B$ in the meantime) is 
\[h\nu = 4d_nE.\]
It is therefore not surprising that essentially all of the systematic errors associated with EDM measurements arise from coupling of the (relatively large) magnetic dipole moment to small changes in the magnetic field that happen to be correlated with the electric field.  

In this section we discuss the available sources of neutrons, as well as the technique, due to Ramsey \cite{ramsey_molec_beams}, by which these extremely precise frequency measurements can be made.  Details of the recently completed room-temperature and the upcoming cryogenic experiments at the Institut Laue-Langevin (ILL) in Grenoble are discussed in the following sections. 

\subsection{Neutron sources}

Early attempts to measure the neutron EDM relied upon beams of cold neutrons from reactors.  This technique eventually became limited by the $\vec{v}\times\vec{E}$ effect: the relativistic transformation of the applied electric field into the neutron's rest frame produces an additional magnetic field, 
\[B' \approx \frac{\vec{v}\times\vec{E}}{c^2},\]
which is proportional to $\vec{E}$ and which therefore can mimic an EDM.  The interaction time $T$ (to which the sensitivity is inversely proportional) is also very short for neutrons in a beam.
Zel$^\prime$dovich \cite{zeldovich59} was the first to propose storing neutrons in material bottles. Fermi \cite{fermi46} had earlier realised that if the neutrons' energies are sufficiently low, their wavelengths become so long that they are no longer able to resolve individual nuclei in the material wall; instead, they sample millions of nuclei simultaneously, and respond to an effective potential -- the Fermi potential $V_F$ -- corresponding to the average bound coherent scattering length.  This Fermi potential may be either positive or negative; thus, for some materials the neutron sees an effective potential barrier and, if its kinetic energy is less than $V_F$, it is able to undergo total (elastic) external reflection.  By surrounding the neutrons with walls of appropriate $V_F$, the neutrons are able to be contained for minutes at a time, thus substantially reducing the $\vec{v}\times\vec{E}$ effect and simultaneously increasing the interaction time by several orders of magnitude. Neutrons in the energy range to which this applies -- up to a few hundred neV, with velocities of a few m/s -- are known as ultra-cold neutrons (UCN) \cite{byrne, golub_UCN_book}.  Bottled UCN are now used routinely in neutron EDM experiments  \cite{baker06, altarev96} (although for a notable exception, see \cite{fedorov01}).

For magnetised materials, the Fermi potential is modified by the addition of a term $-\vec{\mu}\cdot\vec{B}$, and thus depends on the orientation of the neutron's spin relative to the magnetic field.  This property is used to polarize the neutrons by transmission, as the ``wrong'' spin state can be reflected from a magnetised iron foil.

The UCN employed at the Institut Laue-Langevin, Grenoble, are produced from the fission of $^{238}$U nuclei in the enriched-uranium reactor core.  They are allowed to thermalise in a bottle of liquid deuterium, from which they may escape up a vertical guide tube ({\em tube guide verticale}, or TGV).  By the time they reach the top of this tube, they have slowed under gravity to a typical speed of 50 m/s.  In order to slow them further, they are allowed to enter a ``neutron turbine'' \cite{steyerl86}.  The blades of the turbine are receding at $\sim25$ m/s, and thus by bouncing off these blades the neutrons are essentially brought almost to a standstill (although, as a consequence of Liouville's theorem, their angular divergence is much increased).  Neutron densities of the order of 60 cm$^{-3}$ are obtainable with this system.
 
\subsubsection{Superthermal sources}

\begin{figure}[ht]
  \begin{center}
    \resizebox*{0.6\textwidth}{!}{\includegraphics
    [clip=true, viewport = 50 409 540 763]
    {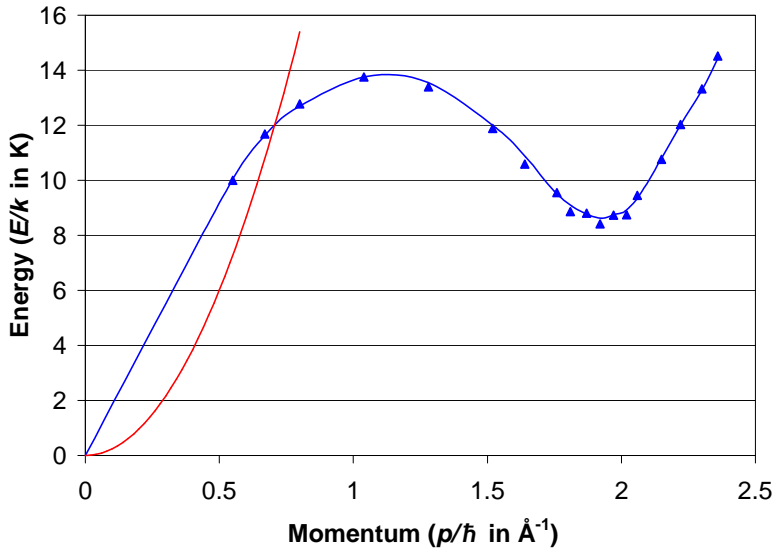}}
    \caption{Energy-momentum dispersion curves for free neutrons (the parabola) and for superfluid $^4$He (with data points from neutron scattering measurements \cite{yarnell59}). 
}
    \label{fig:liq_he_dispersion}
  \end{center}
\end{figure}

Golub and Pendlebury \cite{golub_pendlebury77} in 1977 devised a new technique for the production of UCN, involving the scattering of cold neutrons from nuclei of liquid $^4$He.  The Feynman-Landau phonon-roton dispersion curve for superfluid liquid helium \cite{landau41,landau47,feynman54, yarnell59} is shown in Figure \ref{fig:liq_he_dispersion}, along with that for free neutrons (the latter being parabolic: $E = p^2/2m$).  
At the point where the curves cross, the 8.9 \AA\ neutrons are able to downscatter to the UCN region by emission of a single phonon \cite{cohen57}.  The reverse process -- upscattering -- is completely suppressed by the Boltzmann factor; there are simply no 11 K phonons available to scatter the neutrons within the 0.5 K helium.  As the emitted phonons are able to carry away the excess energy, the restrictions of Liouville's theorem are overcome, and the densities of UCN that are achievable in principle are restricted only by the production rate and the loss rate (the latter being due primarily to the finite neutron lifetime and to losses at the surfaces of the storage vessel).  The neutrons hardly interact at all with one another, and never have the chance to thermalise.  This mechanism is known as ``superthermal'' UCN production because the densities available in this manner are so much higher than from a traditional ``thermal'' neutron source with its Maxwell velocity distribution.  This is the basis of the cryogenic neutron EDM experiment now being commissioned at the ILL, which will be discussed in section \ref{sec:cryoedm}, as well as a similar experiment planned at the SNS in Oak Ridge \cite{sns_edm}.

At PSI, another UCN source is being built, based upon neutrons from a spallation source which are to be cooled in a block of solid deuterium \cite{psi_ucn}.  This will form the basis of another EDM measurement \cite{psi_edm}, which will employ the suitably modified and upgraded equipment from the ILL room-temperature experiment.

\subsection{Ramsey's technique of separated oscillatory fields}

Essentially all attempts to measure the neutron EDM have been based upon the application of Ramsey's technique of separated oscillatory fields \cite{ramsey_molec_beams} to free neutrons, either in beams or stored in bottles.  The technique provides a statistical sensitivity limited entirely by the uncertainty principle: one can only improve the precision by increasing the number of neutrons, their polarization, the storage time or the strength of the applied electric field.

\begin{figure}[ht]
  \begin{center}
    \resizebox*{0.6\textwidth}{!}{\includegraphics
    [clip=true, viewport = 22 126 502 753]
    {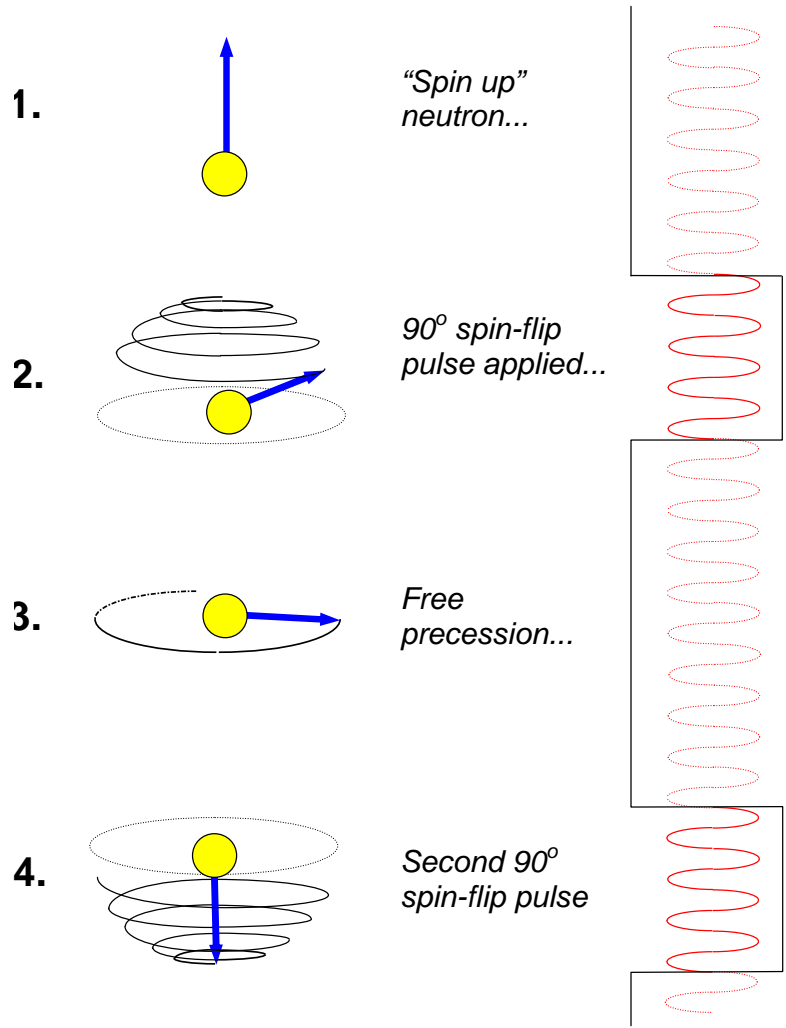}}
    \caption{Ramsey's technique of separated oscillatory fields
}
    \label{fig:ramsey_sequence}
  \end{center}
\end{figure}

Ramsey's technique is perhaps most easily visualised from a semiclassical viewpoint.  Imagine a neutron with its magnetic moment (spin) aligned (anti)parallel to a static homogeneous magnetic field $\vec{B}$ in, say, the vertical direction.  If a small transverse field $\vec{B'}$, rotating at the Larmor frequency, is applied to the system, the neutron in its rest frame will see it as a fixed horizontal field and will begin to precess around it.  This has the effect of making the spin vector spiral around and down as shown in Figure \ref{fig:ramsey_sequence} until, after a time $\tau = 1/(4\gamma B)$ ($\gamma$ being the gyromagnetic ratio), it reaches the horizontal plane.  For obvious reasons this perturbing pulse is referred to as a ``$\pi/2$ pulse".  

The signal generating the perturbing field is then gated off, as shown on the right-hand side of Figure \ref{fig:ramsey_sequence}, and the neutron's spin vector undergoes free precession for some time $T >> \tau$ before a second $\pi/2$ pulse, coherent with the first, is applied.  Between the two $\pi/2$ pulses the system essentially consists of two independent ``clocks'': the source of the $\pi/2$ pulse, which remains active in the background, and the neutron precessing in the magnetic field within its storage volume.  If the frequency of the perturbing field matches precisely the neutron's precession frequency, then the two clocks will stay in phase, and the second pulse will continue to tip the spin vector over until the ``flip'' is completed, as shown in the final diagram of Figure \ref{fig:ramsey_sequence}.  

\begin{figure}[ht]
  \begin{center}
    \resizebox*{0.6\textwidth}{!}{\includegraphics
    [clip=true, viewport = 0 410 548 804]
    {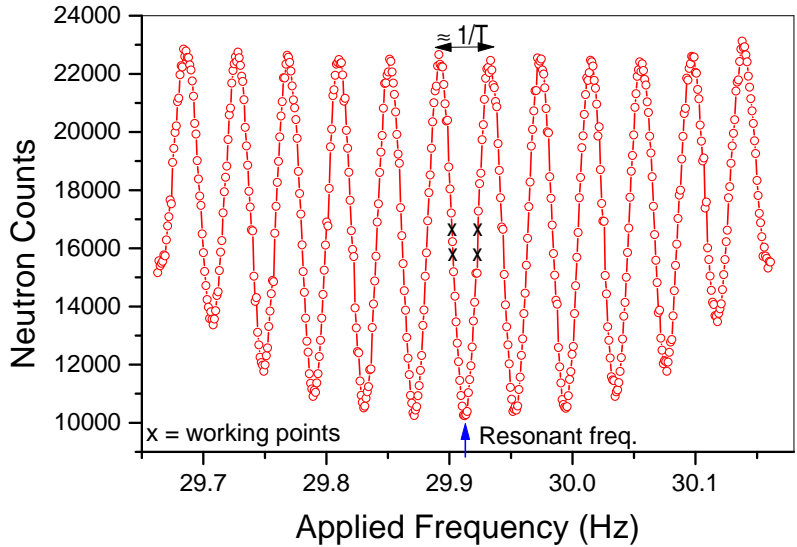}}
    \caption{The number of neutrons remaining with their spins unflipped after application of Ramsey separated oscillatory fields, as a function of the frequency.  From \cite{green98}.
}
    \label{fig:ramseycurve}
  \end{center}
\end{figure}

The beauty of Ramsey's technique is that if the two ``clocks'' have slightly different frequencies, a phase {\em difference} is able to accumulate during the period of free precession.  The second pulse will then be less efficient at tipping the spin vector downwards -- indeed, if the phase difference is 180$^\circ$, then the spin will be tipped back up instead of down.  Figure \ref{fig:ramseycurve} shows the number of spin-up neutrons remaining as a function of the frequency of the applied r.f.\ pulses.  Right at resonance, the maximum number of neutrons have their spins flipped downwards (the fraction being limited by depolarization); 1/$2T$ away from resonance, the acquired phase difference is 180$^\circ$, and the spins are not flipped.

It is interesting to note that Ramsey's idea arose from his teaching undergraduates about the Michelson interferometer\cite{ramsey_private_michelson}.  The analogy is clear.  In fact, the technique (and the resulting graph of Figure \ref{fig:ramseycurve}) is often likened to a two-slit experiment with the slits separated in time rather than space.  

Absolute frequency measurements are not necessary; the EDM depends upon measuring a frequency difference when the electric field direction is reversed.  In order to achieve the maximum sensitivity to such frequency shifts, the experiment operates halfway up the central valley in the regions marked by ``$\times$'' on the curve.  Here, the slope is steepest, so a small frequency shift produces a large change in the number of spin-up neutrons counted.  

\section{ILL room-temperature experiment}

Final results from the room-temperature EDM experiment were published in 2006 \cite{baker06, baker07}.  The brief nature of this review precludes a full and detailed discussion of the experiment, and in particular of the systematic uncertainties associated with it.  A detailed archival paper is, however, in preparation.

\begin{figure}[ht]
  \begin{center}
    \resizebox*{0.6\textwidth}{!}{\includegraphics
    [clip=true, viewport = 69 260 488 763]
    {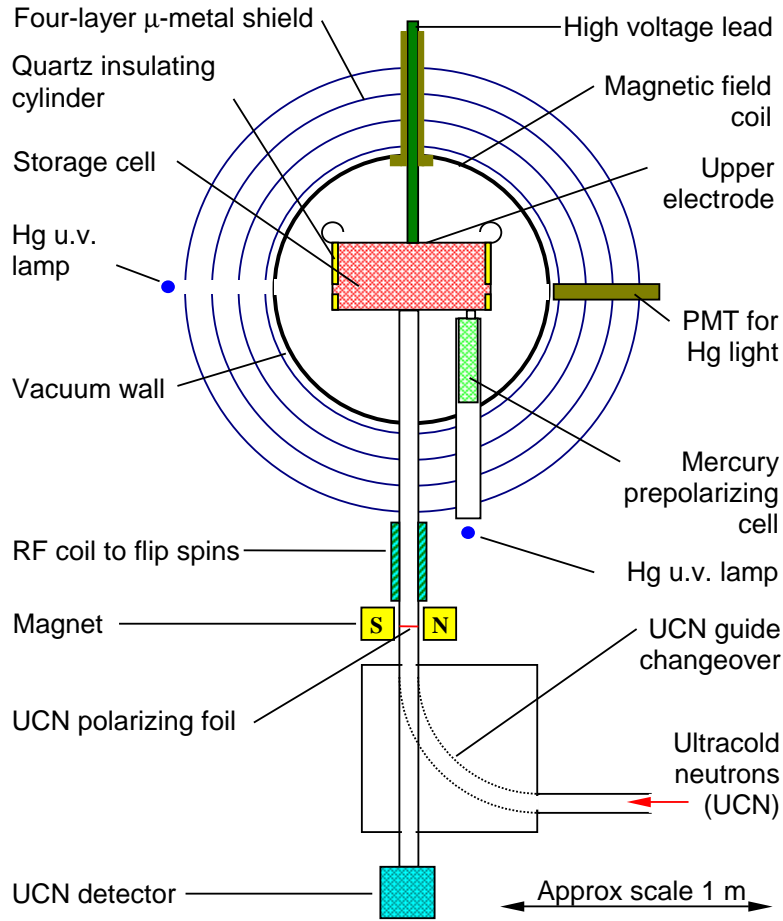}}
    \caption{Apparatus for the room-temperature nEDM experiment at ILL.  From \cite{green98}.
}
    \label{fig:apparatus}
  \end{center}
\end{figure}

A schematic diagram of the room-temperature nEDM experiment at the ILL is shown in Figure \ref{fig:apparatus}.  
Neutrons from the UCN turbine entered the apparatus from the lower right, and flowed upwards (UCN behave much as a diffuse gas) to the polarizer foil.  Those of the correct spin passed through the foil and continued to rise until they reached the storage bottle at the heart of the apparatus.  The bottle was located in a 10 mG (1 $\mu$T) vertical magnetic field; four layers of mu-metal were used to shield out external fields, including that of the Earth.   After a filling period of about 20 seconds, the neutron door was closed, and the Ramsey sequence was applied to the trapped neutrons.  The door was then opened, and the neutrons fell back down to the polarizer, which then acted as an analyser.  Those in the original spin state could pass through and down to a gaseous $^3$He neutron detector (the curved guide tube from the source having been moved aside in favour of a vertical guide leading to the detector).  Neutrons of the ``wrong'' spin state, which bounced off the polarizer, were counted in their turn by employing a fast-passage adiabatic spin flipper to reverse the direction of the spins of the neutrons in the guide tube just above the polarizer: this spin flipper consisted simply of a solenoid wrapped around the guide tube, situated in a region of magnetic field gradient, to which a high-frequency (20 kHz) alternating current was applied.  

Each batch of neutrons takes about four minutes to process.  Between each batch, the applied frequency was cycled between the four marked points on the Ramsey curve of Figure \ref{fig:ramseycurve}.  Periodically (typically hourly) the electric field, which arose simply from the application of 100 kV or so to the lid of the neutron storage container, was reversed. 

\begin{figure}[ht]
  \begin{center}
    {\resizebox*{0.6\textwidth}{!}{\includegraphics
    [clip=true, viewport = 21 85 552 495]
    {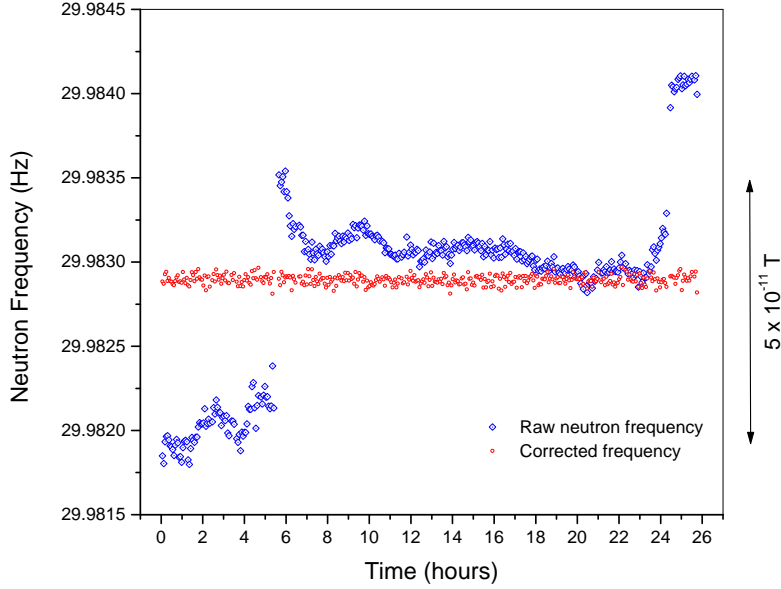}}}
    \caption{Neutron resonant frequency, measured over a 26 hour period, before and after magnetic field drift corrections.  From \cite{green98}.}
    \label{fig:neutronfreq}
  \end{center}
\end{figure}

By fitting successive neutron counts to a sinusoid -- an excellent representation of the central valley of the Ramsey curve of Figure \ref{fig:ramseycurve} -- the resonant frequency was determined. The points marked `Raw neutron frequency' in Figure \ref{fig:neutronfreq} show the behaviour of this frequency over a period of about a day.  No correlation with the hourly reversal of the electric field is apparent.

Figure \ref{fig:neutronfreq} demonstrates clearly the requirement to measure and to control the magnetic environment very precisely.  The four layers of mu-metal provided a radial shielding factor of about 10$^4$, but strong external fields (such as that from the movement of an overhead crane) still had some residual penetration.  Mechanical stresses and thermal expansion of the shields also allowed a degree of leakage of external fields into the inner region.  Batch-by-batch magnetic field variations of the order of 10$^{-12}$ T were common, and sudden jumps an order of magnitude or so larger than this could occur several times per day.  

\subsection{Mercury magnetometer}

\begin{figure}[ht]
  \begin{center}
    \resizebox*{0.35\textwidth}{!}{\includegraphics
    [clip=true, viewport = 47 49 565 743]
    {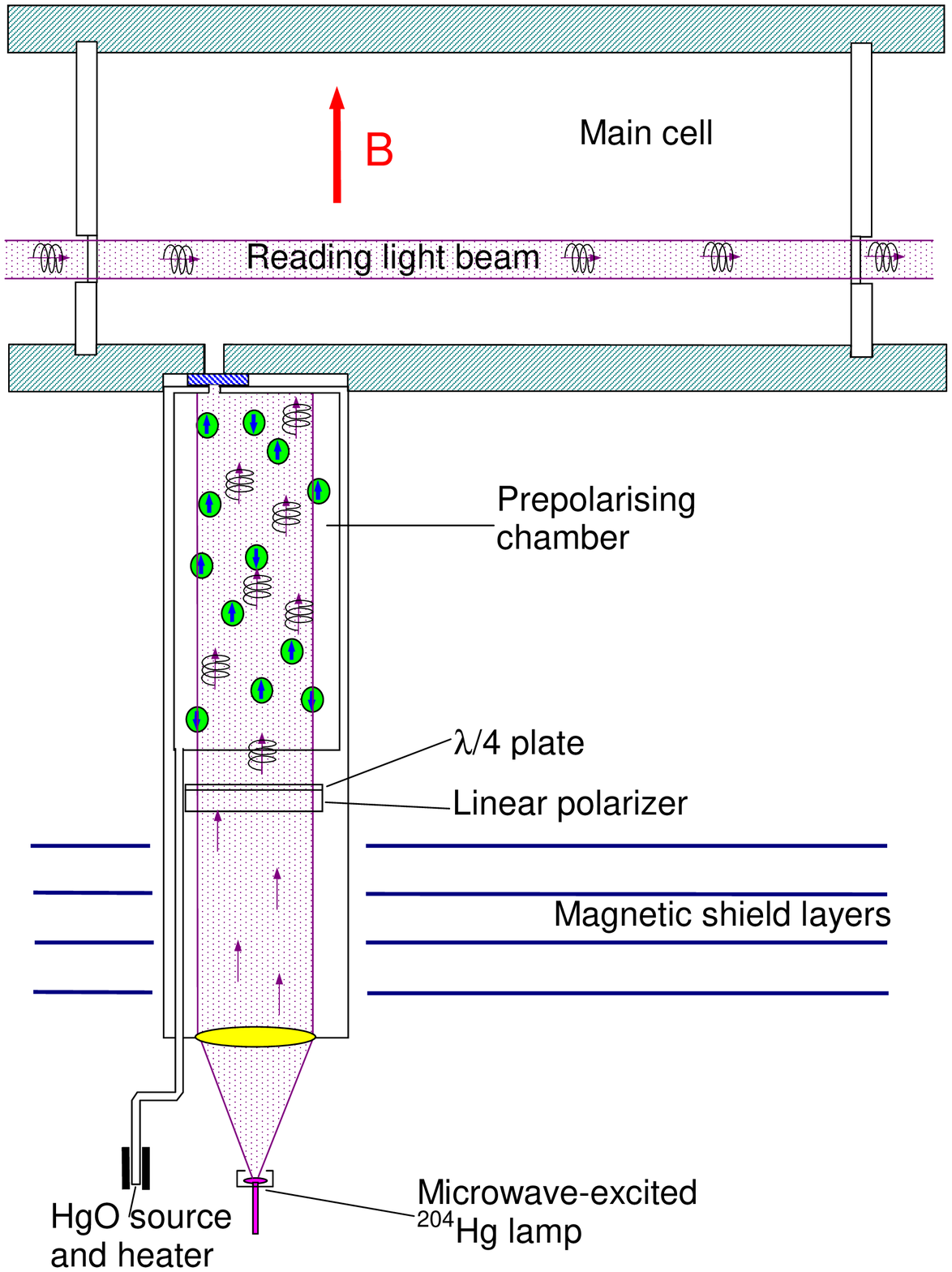}}
    \resizebox*{0.4\textwidth}{!}{\includegraphics
    [clip=true, viewport = 66 315 447 760]
    {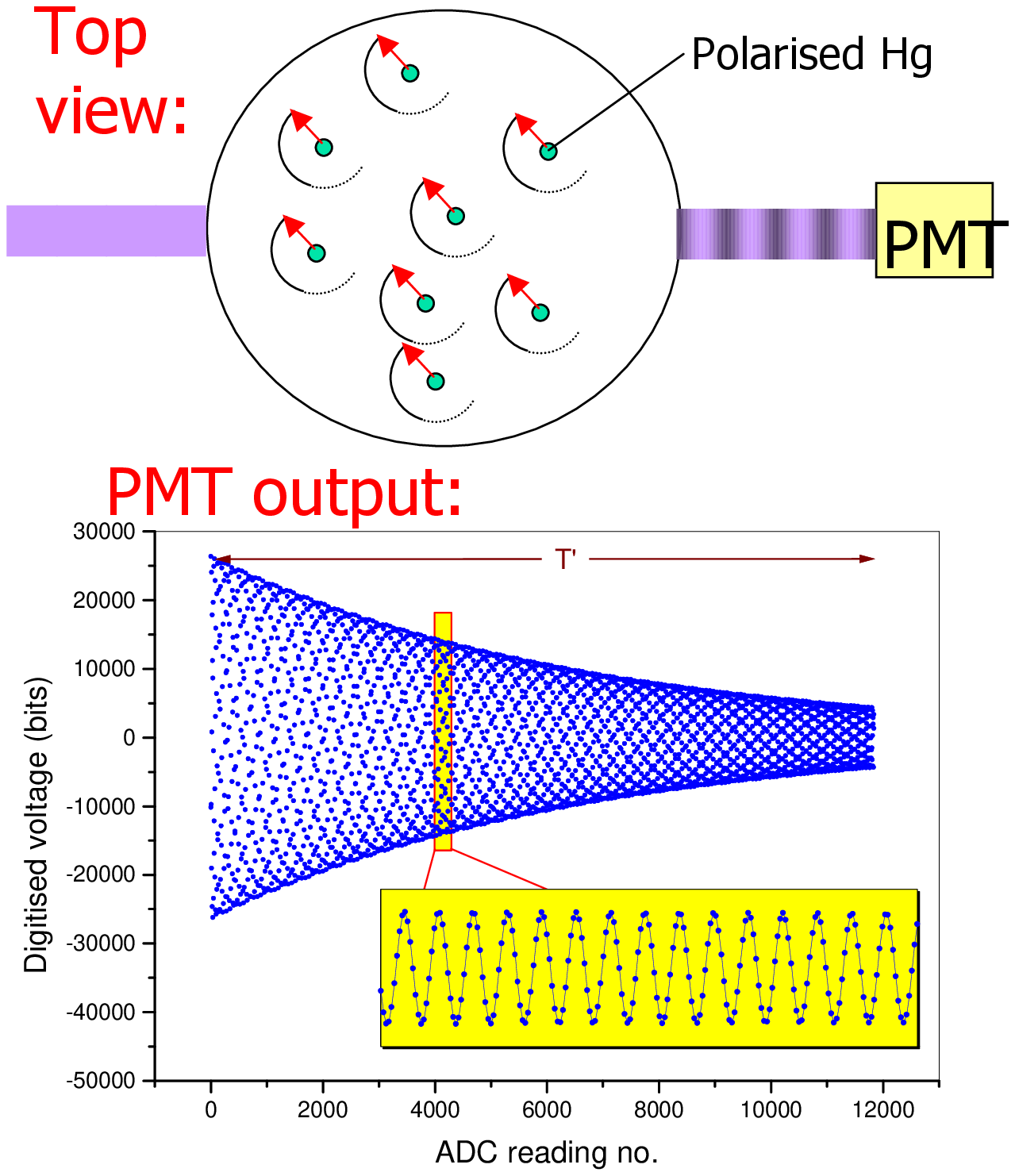}}
    \caption{The structure and operational principle of the mercury magnetometer.  A typical output signal is also shown; the fitted frequency yields the strength of the ambient magnetic field.
}
    \label{fig:merc_mag}
  \end{center}
\end{figure}

Following a suggestion from Ramsey, the magnetic field within the neutron storage volume was monitored by an innovative co-habiting atomic mercury magnetometer \cite{green98}, similar in design to the cell used in the $^{199}$Hg EDM measurement \cite{romalis01}.  Within a small preparation cell, situated just below the neutron storage cell, mercury atoms were polarized by optical pumping.  Just after the neutrons entered their storage vessel, a small door opened briefly to allow the mercury atoms into the same volume.  A Ramsey-type $\pi/2$ pulse was applied to rotate their spins into the horizontal plane, where they were allowed to precess freely.  Circularly polarized light at resonance traversed the cell horizontally; it was absorbed to a greater or lesser extent according to the orientation of the mercury atoms' spin vectors.  Thus, the precession of the atoms throughout the duration of the neutrons' Ramsey measurement sequence imparted an AC component to the intensity of light detected at the far side of the cell.  This signal was digitised and analysed, and the precession frequency that emerged yielded a direct measure of the average strength of the magnetic field within the bottle.  The structure, operational principle and output from the mercury magnetometer are all illustrated within Figure \ref{fig:merc_mag}.  

Figure \ref{fig:neutronfreq} shows the effect of correcting the neutron frequency for the magnetic field drift: the variation is reduced to a small scatter about the mean, dominated entirely by neutron counting statistics.  

The compensation provided by the magnetometer was not perfect. The neutrons were of such low energy that they populated preferentially the lower portion of the bottle; their centre of mass lay about 2.8 mm below that of the mercury atoms.  Any vertical magnetic field gradient thus resulted in a slightly different average field being seen by the two systems.  However, the systematic uncertainty in the measurement is reduced by more than an order of magnitude.

\subsection{Sensitivity}

Extracting a value for the EDM is, in principle, simple: A straight-line fit to the corrected frequency as a function of the applied electric field $E$ yields a gradient that is directly proportional to the EDM.  The statistical uncertainty obtained is 
\[
\sigma _{d_{n}}\approx \frac{\hbar }{2\alpha ET\sqrt{N}},  \label{stat error}
\]
where $N$ is the total number of neutrons counted, the Ramsey period $T >> t$, and $\alpha $ is the
visibility (averaged over the two spin states) of the central resonance fringe: 
\[
\alpha_\uparrow =\frac{\left( N_{\uparrow \max }-N_{\uparrow \min }\right) }{\left(
N_{\uparrow \max }+N_{\uparrow \min }\right) },
\]
with a similar expression for $\alpha_{\downarrow }$.  The visibility is effectively a measure of the polarization that can be achieved and maintained throughout the measurement period.  The experiment ultimately reached a statistical sensitivity of $\sigma_{d_n} \approx 1.5\times10^{-26}\ e$cm.

\section{Systematic uncertainties}

As noted above, the systematic uncertainties in this experiment principally arise from changes in the magnetic field, as seen by the neutrons, that are correlated with the electric field (or, more precisely, with the electric field times the sign of $\vec{B_0}$).  Changes in $B$ that are uniform throughout the volume are compensated by the mercury, but any change in the vertical gradient cannot be completely compensated.  There are several possible causes of systematic uncertainty; here we outline briefly those that make the most significant contributions.

\begin{itemize}

\item{} Geometric phase effect.  This is the leading source of systematic error in the room-temperature experiment, and will be discussed in Section \ref{sec:GP_effect}.

\item{} If the $^{199}$Hg were to have a sizeable EDM of its own, the magnetic field compensation would yield a false apparent neutron EDM (possibly masking a true neutron EDM) of magnitude $d_{Hg}\cdot\gamma_n/\gamma_{Hg}$ (where the $\gamma$ factors are the gyromagnetic ratios), but this is not the case here as the EDM of $^{199}$Hg has been shown to be insignificant on this scale \cite{romalis01}.  

\item{} Residual fields from external influences, such as the movement of an overhead crane.

\item{} Currents in the bottle.  If any leakage current follows an azimuthal path around the bottle as it flows from one electrode to the other, it will generate an axial magnetic field.  

\item{} Sparks, leading to changes in the residual magnetisation of the shielding.  

\item{} Oscillating fields, perhaps due to ripple on the high-voltage line, causing frequency shifts through Ramsey-Bloch-Siegert \cite{bloch40,ramsey55} type effects.

\item{} $\vec{v}\times\vec{E}$ fields, where the velocity in question arises from either average centre-of-mass motion or from rotation of the neutrons within the bottle.  The direct influence of such fields is generally very small for trapped ultracold neutrons.  There is also a second-order $\vec{v}\times\vec{E}$ effect, due to the slightly enhanced total magnitude of the magnetic field.  This gives a signal proportional to $E^2$ rather than to $E$, but was also a negligible effect in this experiment.

\item{} Light-induced shifts in the Hg resonant frequency.  This affects the result in two ways: First, by shifting the measured frequency, it alters the geometric-phase correction.  Second, if the amplitude of the light was correlated in some way with the electric field, the frequency shift would reflect this and produce a signal directly.  

\item{} Mechanical movement of an electrode due to electrostatic forces.  This effect is minimised by having sturdy electrodes held rigidly in place.
 
\end{itemize}

\subsection{Geometric phase effect}
\label{sec:GP_effect}

\begin{figure}[ht]
  \begin{center}
    \resizebox*{0.35\textwidth}{!}{\includegraphics
    [clip=true, viewport = 82 500 339 774]
    {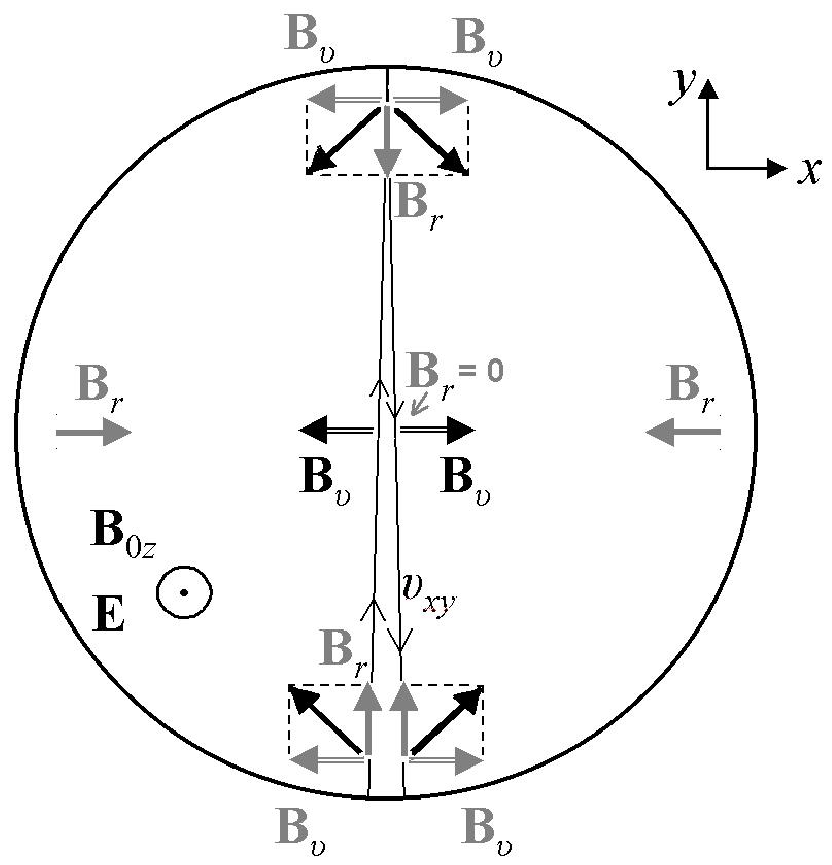}}
    \resizebox*{0.5\textwidth}{!}{\includegraphics
    [clip=true, viewport = 53 65  530 390]
    {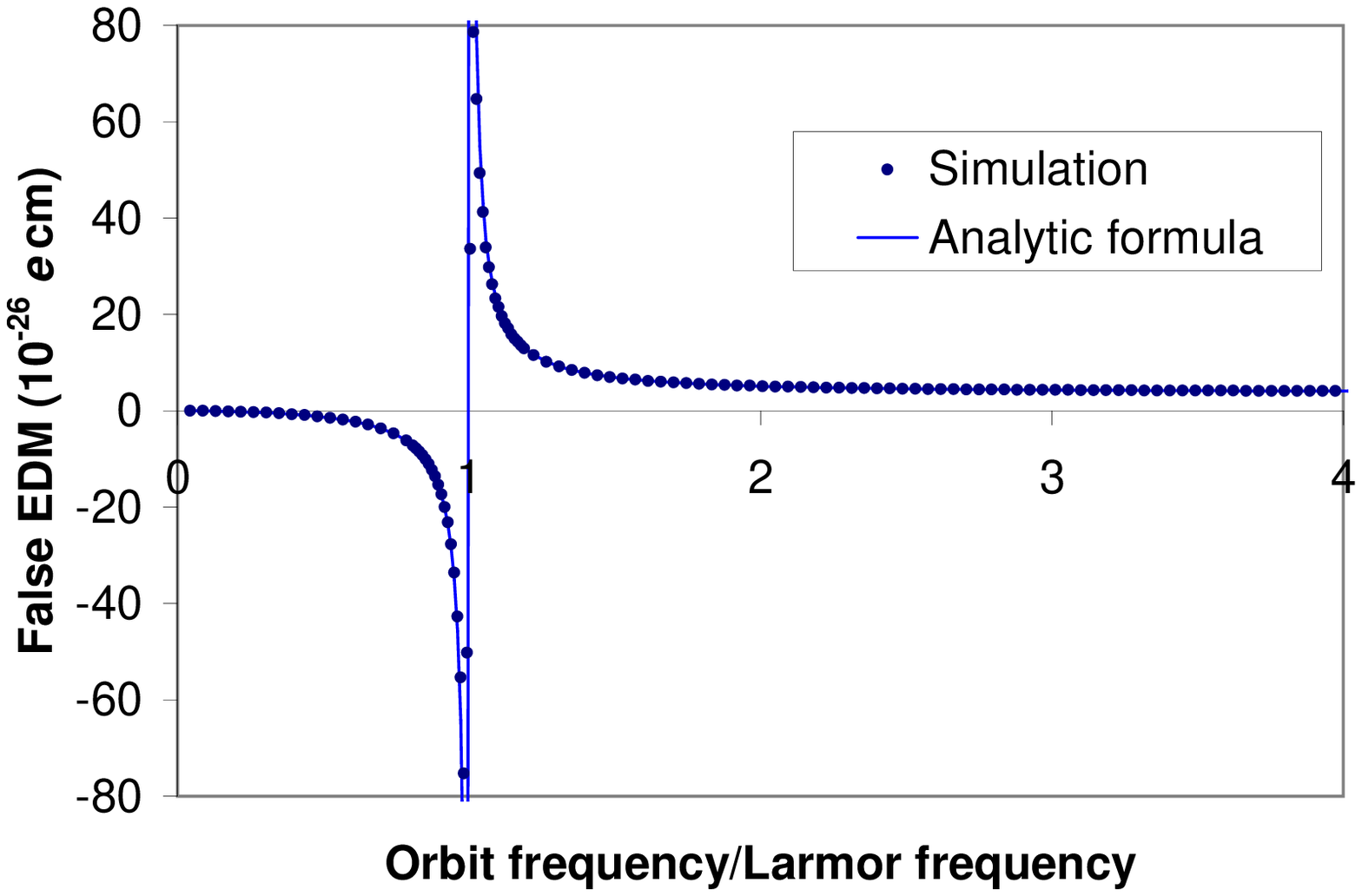}}
    \caption{The origin of the geometric-phase induced false EDM effect, and its magnitude as a function of orbit frequency for a 1 nT/m vertical magnetic field gradient.  Both of these figures are from \cite{pendlebury04}; copyright American Physical Society.
}
    \label{fig:geophase_effect2}
  \end{center}
\end{figure}

The leading systematic uncertainty in the room-temperature EDM experiment arises from a subtle and beautiful conspiracy, only recently discovered \cite{pendlebury04}, between the $\vec{v}\times\vec{E}$ effect and a vertical gradient in the magnetic field.  The situation is easiest to understand in the case of an azimuthally symmetric field with a vertical gradient  $\partial B/\partial z$, i.e., a slightly trumpet-shaped field.  Because $\vec{\nabla}\cdot\vec{B} = 0$, field lines either enter or emerge from the side walls of the cell, giving a radial field of strength that is proportional to the radius $r$:
\[B_r = -\frac{r}{2}\cdot\frac{\partial B}{\partial z}.\]
Consider now a particle crossing the storage cell close to its diameter, as shown in Figure \ref{fig:geophase_effect2}.  
At the start of its trajectory, just left of centre at the bottom, it is subject to the radial field $B_r$ as well as to the sideways $\vec{v}\times\vec{E}$ component, yielding a diagonal resultant.  As it moves up the bottle, the $B_r$ component shrinks and then reverses direction, causing a smooth rotation of this additional effective $B$ field.  Eventually, it reaches the far end of the bottle, is reflected from the wall and begins its trajectory back.  However, the $\vec{v}\times\vec{E}$ component then faces in the opposite direction, with the net effect that the additional effective field component {\em continues to rotate in the same direction}.  The particle thus sees a rotating field in the $x$-$y$ plane, which, through the Ramsey-Bloch-Siegert mechanism, `pulls' its resonant frequency away from the central value.  The shift in frequency is proportional to $E$, and it therefore mimics an EDM signal.  This false EDM  is shown in Figure \ref{fig:geophase_effect2} as a function of the diametric orbit frequency $v/(4R)$ normalised to the Larmor frequency.

For the neutrons, moving at just a few m/s in a bottle of diameter 0.5 m, the orbit frequency is rather low compared to the Larmor frequency.  In this limit, the induced false EDM is small, negative,  increases quadratically with velocity, and, as it turns out, is independent of the trajectory within the bottle.   

\begin{figure}[ht]
  \begin{center}
    \resizebox*{0.8\textwidth}{!}{\includegraphics
    [clip=true, viewport = 50 205 542 774]
    {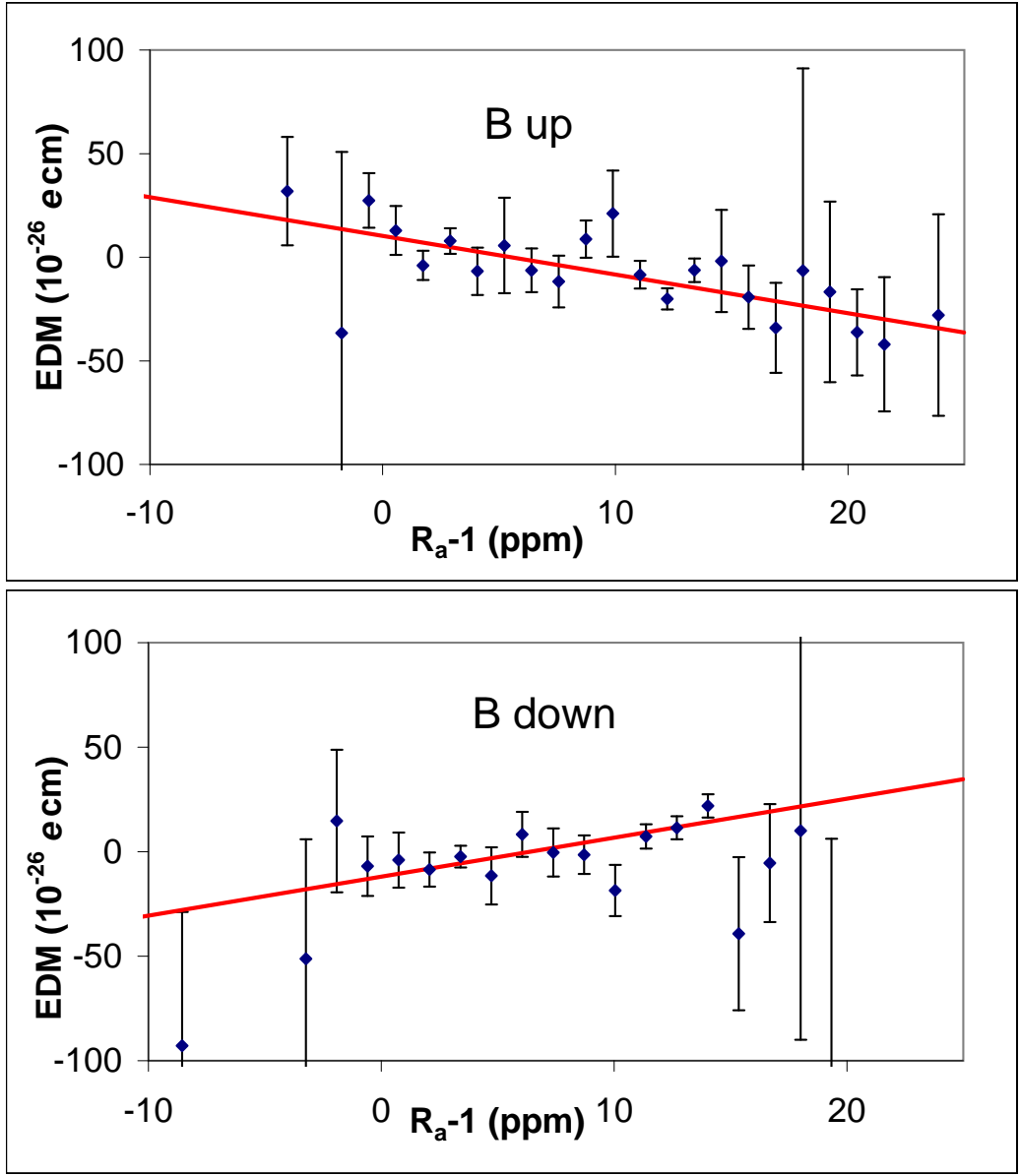}}
     \caption{The measured neutron EDM as a function of the neutron-to-mercury frequency ratio.  From \cite{baker06}; copyright American Physical Society.
}
    \label{fig:edm_vs_R}  
  \end{center}
\end{figure}

The mercury atoms used for magnetic field compensation are in the opposite regime.  For them the shift is relatively large and is independent of velocity (although in their case there is some dependence upon the trajectory within the bottle).  Thus, in the presence of a vertical field gradient, the use of the mercury magnetometer to compensate for drifts in the $B_0$ field actually {\em introduces} a false EDM effect that is proportional to $\partial B/\partial z$.  As is shown in Fig.\ \ref{fig:edm_vs_R}, the effect is clearly visible within the data.  For uniform vertical field gradients, however, this geometric-phase effect can be entirely eliminated by using the crossing point of the lines in Fig.\ \ref{fig:edm_vs_R}, once allowance has been made for the Earth's rotation, which shifts the frequency ratio measurements a little lower (higher) when the $\vec{B}$ field is upwards (downwards) \cite{lamoreaux07, baker07}.  (This effect\footnote{The importance of considering the Earth's rotation in EDM measurements may have been appreciated first by J. Miller, who documented it in a number of $g-2$ collaboration internal notes and brought it to the attention of the larger EDM community during discussions at the 2006 Lepton Moments conference.}   is analogous to Foucault's pendulum: the neutrons and mercury atoms precess freely while the apparatus rotates slowly beneath them.)

The siutation is further complicated by an additional effect due to horizontal magnetic fields $B_x$, $B_y$ with finite $\partial B_x/\partial y$ and/or $\partial B_y/\partial x$ but with ($\partial B_x/\partial x + \partial B_y/\partial y) = 0= -\partial B_z/\partial z$ (e.g., a quadrupole aligned with z, with $B_x = qy, B_y = qx$).  These fields -- henceforward referred to for simplicity as horizontal quadrupolar fields -- do not of themselves induce false EDM signals, but they shift the ratio of frequencies of the neutrons and mercury atoms.  The underlying mechanism is that the slow-moving neutrons are able to follow the field variations adiabatically as they traverse the bottle, and they therefore see a total $B_0$ field of slightly enhanced amplitude and slightly varying direction.  The mercury atoms are moving so fast that they simply average out the small horizontal components.  If these horizontal quadrupolar fields are the same for both polarities of the magnetic field, the lines of Fig.\ \ref{fig:edm_vs_R} are shifted sideways, and there is no net change to the measured (crossing point) EDM.  If this is not the case, however, a sideways shift of one line relative to the other does shift the crossing point up or down.  Local magnetic dipole fields can also enhance the geometric-phase shift, beyond the expectation arising from the volume-averaged magnetic field gradient \cite{harris06}.  

The details of the analysis are beyond the scope of this review, but essentially one must rely upon independent measurements of the frequency ratios (for each of the field directions) at which the volume-averaged magnetic field gradient is zero.  The procedure is outlined in  \cite{harris06}, and will be addressed further in an archival paper.

\section{Cryogenic neutron EDM experiment}
\label{sec:cryoedm}

The prediction that a high flux of neutrons could be produced from a superthermal source was first made in 1977 \cite{golub_pendlebury77}.  During the intervening period, a number of attempts were made to measure the production rate \cite{yoshiki92}, but all were severely hampered by the necessity of bringing the neutrons out of the superfluid helium volume to the detectors, which, being gaseous, were at room temperature.  Both the windows used, and the contaminants that invariably froze onto them, proved to be very effective neutron absorbers.  In 2001, however, the Sussex/RAL EDM group developed solid-state neutron detectors that could be placed within the superfluid helium itself \cite{baker03a}.  The detectors were coated with a layer of $^6$Li, and the neutrons were thus able to be detected via the reaction
\begin{center}n + $^6$Li $\rightarrow \alpha + ^3$H.\end{center}
With this technology in place, the production rate could be measured, and was found to be in good agreement with the theoretical prediction \cite{baker03b}.  

\begin{figure}[ht]
  \begin{center}
    \resizebox*{0.8\textwidth}{!}{\includegraphics
    [clip=true, viewport = 30 430 565 780]
    {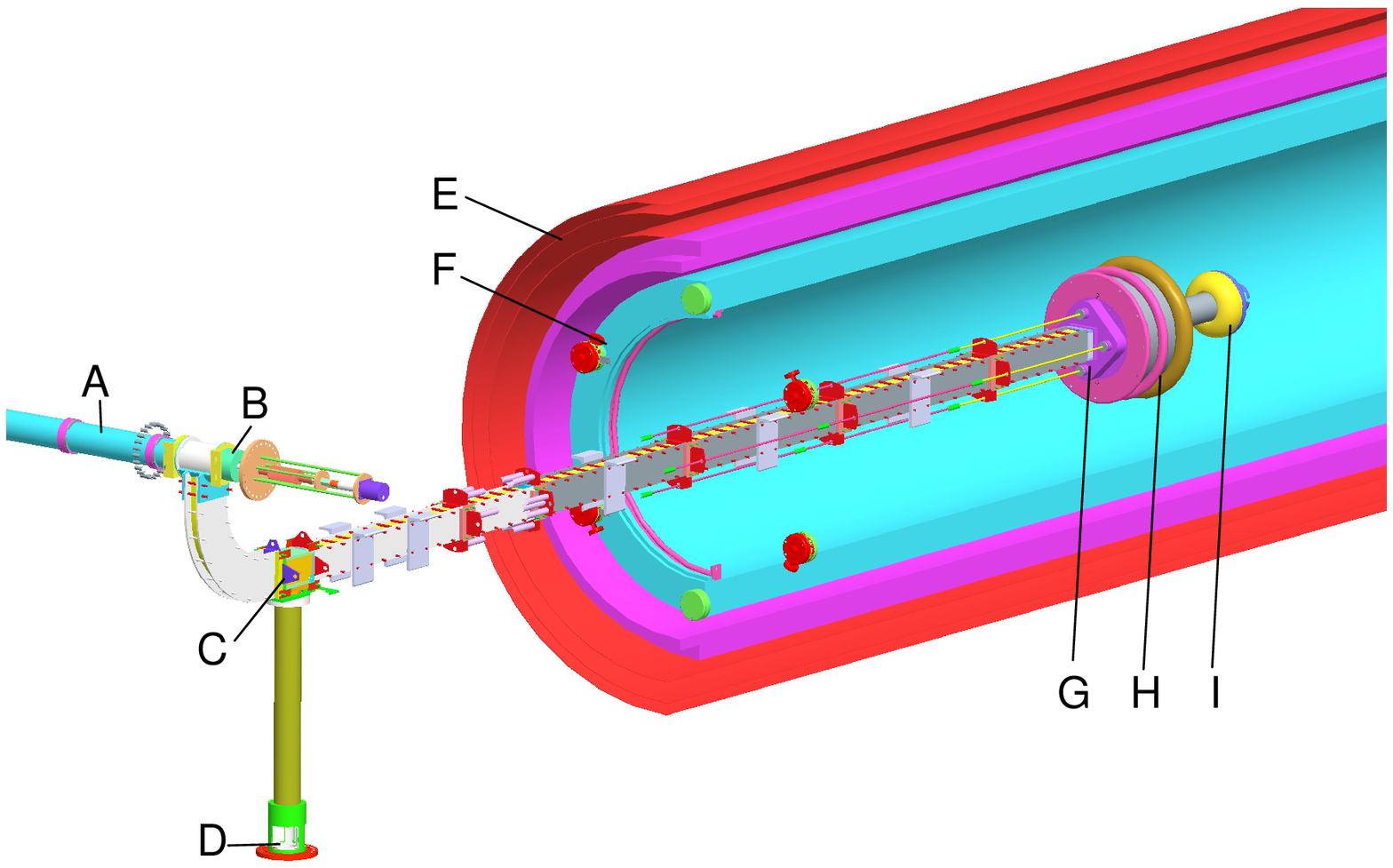}}
    \caption{A schematic diagram of the cryogenic neutron EDM apparatus being developed by the Sussex/RAL/ILL/Kure/Oxford CryoEDM collaboration.  Key: [A] UCN source volume [B] Valve and beam stop [C] Flap valve, to direct neutrons either from source to Ramsey cells or from Ramsey cells to detector [D] Detector [E] three-layer mu-metal shielding [F] Helium jacket, containing superconducting lead shield and superconducting solenoid [G] SQUID pickup coil location [H] Ramsey cells [I] High-voltage feedthrough; extended high-voltage feed line not shown.}
    \label{fig:cryoedm_eqpt}
  \end{center}
\end{figure}

Figure \ref{fig:cryoedm_eqpt} shows schematically the apparatus that is being developed by the same group -- joined in the meantime by H.\ Yoshiki of the University of Kure in Japan, who provided much of the equipment \cite{yoshiki96}, and by the University of Oxford --  for the cryogenic EDM experiment \cite{cryoedm_proposal} that is anticipated to take place at the ILL.  

A polarizing monochromator is used to extract the 8.9 \AA  \ neutrons from the ILL's H53 cold-neutron beamline.  Cooling Towers I and II provide liquid helium for, respectively, the neutron-containing volumes and for the superconducting shield/solenoid arrangement.  Not shown is an extended neutron production volume and a 90$^\circ$ bend between Towers I and II so that the cold neutron beam does not enter the measurement volume.  The measurement process will follow the time-honoured fashion of allowing the neutrons into the Ramsey chamber in batches, carrying out the usual spin manipulations to determine the Larmor precession frequency, and then detecting and counting the number of neutrons in each spin state. It is planned that the Ramsey chamber at the heart of the apparatus will eventually consist of four cells: the two central cells, separated by a high-voltage electrode, will have oppositely directed electric fields; the outer cells, with grounded electrodes at either end, will act as neutron magnetometers.  (Initially, there will be just two cells: one subject to an electric field, the other in which both electrodes are grounded.)  Although a cohabiting magnetometer has been very successful in the room-temperature experiment, the geometric-phase induced systematic false EDM effect has led to a decision not to employ a similar technique in this experiment:  The magnetic field will be monitored by SQUIDs.  A 400 kV Cockcroft-Walton stack will provide high voltage, input from the right-hand side of the figure. 

At the time of writing (autumn 2007) much of the construction is complete, and commissioning is underway.  Within a relatively short time, a sensitivity of 10$^{-27}\ e$cm should be achieved.  It is then hoped that the experiment may move to a dedicated, high-flux beamline.  A three- to five-year running period should then enable the statistical sensitivity to be brought down to the region of a few times 10$^{-28}\ e$cm, thus either verifying or testing to the limit many popular models -- such as supersymmetry -- that attempt to explain the origins of CP violation.

\section{Conclusions and Future Outlook}

Neutron EDM experiments provide a unique and sensitive probe to physics beyond the Standard Model.  Models that predict a sufficient degree of CP violation to account for the observed baryon asymmetry of the Universe typically also predict the existence of a neutron EDM at, or extremely close to, the existing experimental limits: as such, these ever-tightening limits provide severe constraints on such models, including, for example, Supersymmetry.

The most sensitive experiment to date has a statistical uncertainty that is approaching the level of 10$^{-26}\ e$cm.  While it may be possible to gain perhaps a factor of three or four improvement in sensitivity by modifying the apparatus in question and situating it at a more intense neutron source \cite{psi_edm}, or else by building a new apparatus with multiple chambers to improve the cancellation of false EDM signals \cite{serebrov_edm}, the possibility of a truly dramatic breakthrough appears to require a cryogenic experiment using the so-called superthermal production method.  One such experiment, highlighted in this article, is currently under development; others have been proposed, for example by a collaboration in the USA \cite{cooper} that proposes to use a $^3$He cohabiting magnetometer \cite{golub94} in order ultimately to achieve a sensitivity of about 10$^{-29}\ e$cm.   

We are fortunate to be working in the field at this very exciting time: if Supersymmetry is correct, then it is quite likely that within the next five to ten years this half-century long search for a finite electric dipole moment of the neutron will finally reach its goal.

\section*{Acknowledgements}

The author would like to acknowledge his enormous debt of gratitude to his colleagues and collaborators on the neutron EDM experiments, both room-temperature and cryogenic.  Particular thanks are due to Mike Pendlebury and to Keith Green, who for many years have provided tremendous intellectual and organisational leadership for these projects; and to Norman Ramsey whose brilliance and insight started the journey more than fifty years ago, and whose continuing support, interest and involvement has been so fruitful.  The author is also extremely grateful to Steve Abel for clarifications, suggestions and helpful discussions relating to the theory.


\begin{thebibliography}{10}
\expandafter\ifx\csname natexlab\endcsname\relax\def\natexlab#1{#1}\fi
\expandafter\ifx\csname bibnamefont\endcsname\relax
  \def\bibnamefont#1{#1}\fi
\expandafter\ifx\csname bibfnamefont\endcsname\relax
  \def\bibfnamefont#1{#1}\fi
\expandafter\ifx\csname citenamefont\endcsname\relax
  \def\citenamefont#1{#1}\fi
\expandafter\ifx\csname url\endcsname\relax
  \def\url#1{\texttt{#1}}\fi
\expandafter\ifx\csname urlprefix\endcsname\relax\def\urlprefix{URL }\fi
\providecommand{\bibinfo}[2]{#2}
\providecommand{\eprint}[2][]{\url{#2}}

\bibitem[{\citenamefont{Smith et~al.}(1957)\citenamefont{Smith, Purcell, and
  Ramsey}}]{smith57}
\bibinfo{author}{\bibfnamefont{J.}~\bibnamefont{Smith}},
  \bibinfo{author}{\bibfnamefont{E.}~\bibnamefont{Purcell}}, \bibnamefont{and}
  \bibinfo{author}{\bibfnamefont{N.}~\bibnamefont{Ramsey}},
  \bibinfo{journal}{Phys.\ Rev.} \textbf{\bibinfo{volume}{108}},
  \bibinfo{pages}{120} (\bibinfo{year}{1957}).

\bibitem[{\citenamefont{Purcell and Ramsey}(1950)}]{purcell50}
\bibinfo{author}{\bibfnamefont{E.}~\bibnamefont{Purcell}} \bibnamefont{and}
  \bibinfo{author}{\bibfnamefont{N.}~\bibnamefont{Ramsey}},
  \bibinfo{journal}{Phys.\ Rev.} \textbf{\bibinfo{volume}{78}},
  \bibinfo{pages}{807} (\bibinfo{year}{1950}).

\bibitem[{\citenamefont{Wu et~al.}(1957)}]{wu57}
\bibinfo{author}{\bibfnamefont{C.}~\bibnamefont{Wu}} \bibnamefont{et~al.},
  \bibinfo{journal}{Phys.\ Rev.} \textbf{\bibinfo{volume}{105}},
  \bibinfo{pages}{1413} (\bibinfo{year}{1957}).

\bibitem[{\citenamefont{Pendlebury and Hinds}(2000)}]{pendlebury_hinds00}
\bibinfo{author}{\bibfnamefont{J.}~\bibnamefont{Pendlebury}} \bibnamefont{and}
  \bibinfo{author}{\bibfnamefont{E.}~\bibnamefont{Hinds}},
  \bibinfo{journal}{Nucl.\ Instr.\ Meth.\ A} \textbf{\bibinfo{volume}{440}},
  \bibinfo{pages}{471} (\bibinfo{year}{2000}).

\bibitem[{\citenamefont{Baker et~al.}(2006)}]{baker06}
\bibinfo{author}{\bibfnamefont{C.}~\bibnamefont{Baker}} \bibnamefont{et~al.},
  \bibinfo{journal}{Phys.\ Rev.\ Lett.} \textbf{\bibinfo{volume}{97}},
  \bibinfo{pages}{131801} (\bibinfo{year}{2006}).

\bibitem[{\citenamefont{Baker et~al.}(2007)}]{baker07}
\bibinfo{author}{\bibfnamefont{C.}~\bibnamefont{Baker}} \bibnamefont{et~al.},
  \bibinfo{journal}{Phys.\ Rev.\ Lett.} \textbf{\bibinfo{volume}{98}},
  \bibinfo{pages}{149102} (\bibinfo{year}{2007}).

\bibitem[{\citenamefont{Trodden}(1999)}]{trodden99}
\bibinfo{author}{\bibfnamefont{M.}~\bibnamefont{Trodden}},
  \bibinfo{journal}{Rev.\ Mod.\ Phys.} \textbf{\bibinfo{volume}{71}},
  \bibinfo{pages}{1463} (\bibinfo{year}{1999}).

\bibitem[{\citenamefont{Ellis et~al.}(1981{\natexlab{a}})\citenamefont{Ellis,
  Gaillard, Nanopoulos, and Rudaz}}]{ellis81a}
\bibinfo{author}{\bibfnamefont{J.}~\bibnamefont{Ellis}},
  \bibinfo{author}{\bibfnamefont{M.}~\bibnamefont{Gaillard}},
  \bibinfo{author}{\bibfnamefont{D.}~\bibnamefont{Nanopoulos}},
  \bibnamefont{and} \bibinfo{author}{\bibfnamefont{S.}~\bibnamefont{Rudaz}},
  \bibinfo{journal}{Phys.\ Lett.\ B} \textbf{\bibinfo{volume}{99}},
  \bibinfo{pages}{101} (\bibinfo{year}{1981}{\natexlab{a}}).

\bibitem[{\citenamefont{Ellis et~al.}(1981{\natexlab{b}})\citenamefont{Ellis,
  Gaillard, Nanopoulos, and Rudaz}}]{ellis81b}
\bibinfo{author}{\bibfnamefont{J.}~\bibnamefont{Ellis}},
  \bibinfo{author}{\bibfnamefont{M.}~\bibnamefont{Gaillard}},
  \bibinfo{author}{\bibfnamefont{D.}~\bibnamefont{Nanopoulos}},
  \bibnamefont{and} \bibinfo{author}{\bibfnamefont{S.}~\bibnamefont{Rudaz}},
  \bibinfo{journal}{Nature} \textbf{\bibinfo{volume}{293}}, \bibinfo{pages}{41}
  (\bibinfo{year}{1981}{\natexlab{b}}).

\bibitem[{\citenamefont{Perkins}(1987)}]{perkins}
\bibinfo{author}{\bibfnamefont{D.}~\bibnamefont{Perkins}},
  \emph{\bibinfo{title}{Introduction to High-Energy Physics, 3rd Ed.}}
  (\bibinfo{publisher}{Addison-Wesley}, \bibinfo{year}{1987}).

\bibitem[{\citenamefont{Christenson et~al.}(1964)\citenamefont{Christenson,
  Cronin, Fitch, and Turlay}}]{christenson64}
\bibinfo{author}{\bibfnamefont{J.}~\bibnamefont{Christenson}},
  \bibinfo{author}{\bibfnamefont{J.}~\bibnamefont{Cronin}},
  \bibinfo{author}{\bibfnamefont{V.}~\bibnamefont{Fitch}}, \bibnamefont{and}
  \bibinfo{author}{\bibfnamefont{R.}~\bibnamefont{Turlay}},
  \bibinfo{journal}{Phys.\ Rev.\ Lett.} \textbf{\bibinfo{volume}{13}},
  \bibinfo{pages}{138} (\bibinfo{year}{1964}).

\bibitem[{\citenamefont{Ellis}(1989)}]{ellis89}
\bibinfo{author}{\bibfnamefont{J.}~\bibnamefont{Ellis}},
  \bibinfo{journal}{Nucl.\ Instr.\ Meth.\ A} \textbf{\bibinfo{volume}{284}},
  \bibinfo{pages}{33} (\bibinfo{year}{1989}).

\bibitem[{\citenamefont{Pospelov and Ritz}(2001)}]{pospelov01}
\bibinfo{author}{\bibfnamefont{M.}~\bibnamefont{Pospelov}} \bibnamefont{and}
  \bibinfo{author}{\bibfnamefont{A.}~\bibnamefont{Ritz}},
  \bibinfo{journal}{Phys.\ Rev.\ D} \textbf{\bibinfo{volume}{63}},
  \bibinfo{pages}{073015} (\bibinfo{year}{2001}).

\bibitem[{\citenamefont{Ellis and Flores}(1996)}]{ellis96}
\bibinfo{author}{\bibfnamefont{J.}~\bibnamefont{Ellis}} \bibnamefont{and}
  \bibinfo{author}{\bibfnamefont{R.}~\bibnamefont{Flores}},
  \bibinfo{journal}{Phys.\ Lett.\ B} \textbf{\bibinfo{volume}{377}},
  \bibinfo{pages}{83} (\bibinfo{year}{1996}).

\bibitem[{\citenamefont{He et~al.}(1988)\citenamefont{He, McKellar, and
  Pakvasa}}]{he88}
\bibinfo{author}{\bibfnamefont{X.-G.} \bibnamefont{He}},
  \bibinfo{author}{\bibfnamefont{B.}~\bibnamefont{McKellar}}, \bibnamefont{and}
  \bibinfo{author}{\bibfnamefont{S.}~\bibnamefont{Pakvasa}},
  \bibinfo{journal}{Phys.\ Rev.\ Lett.} \textbf{\bibinfo{volume}{61}},
  \bibinfo{pages}{1267} (\bibinfo{year}{1988}).

\bibitem[{\citenamefont{Dar}(2000)}]{dar00}
\bibinfo{author}{\bibfnamefont{S.}~\bibnamefont{Dar}},
  \bibinfo{journal}{hep-ph/0008248}  (\bibinfo{year}{2000}).

\bibitem[{\citenamefont{Donoghue et~al.}(1992)\citenamefont{Donoghue, Golowich,
  and Holstein}}]{donoghue}
\bibinfo{author}{\bibfnamefont{J.}~\bibnamefont{Donoghue}},
  \bibinfo{author}{\bibfnamefont{E.}~\bibnamefont{Golowich}}, \bibnamefont{and}
  \bibinfo{author}{\bibfnamefont{B.}~\bibnamefont{Holstein}},
  \emph{\bibinfo{title}{Dynamics of the Standard Model}}
  (\bibinfo{publisher}{Cambridge}, \bibinfo{year}{1992}).

\bibitem[{\citenamefont{Aubert et~al.}(2004)}]{babar04}
\bibinfo{author}{\bibfnamefont{B.}~\bibnamefont{Aubert}} \bibnamefont{et~al.},
  \bibinfo{journal}{Phys.\ Rev.\ Lett.} \textbf{\bibinfo{volume}{93}},
  \bibinfo{pages}{131801} (\bibinfo{year}{2004}),
  \bibinfo{note}{hep-ex/0407057}.

\bibitem[{\citenamefont{Abe et~al.}(2004)}]{belle04}
\bibinfo{author}{\bibfnamefont{K.}~\bibnamefont{Abe}} \bibnamefont{et~al.},
  \bibinfo{journal}{Phys.\ Rev.\ Lett.} \textbf{\bibinfo{volume}{93}},
  \bibinfo{pages}{021601} (\bibinfo{year}{2004}).

\bibitem[{\citenamefont{Romalis et~al.}(2001)\citenamefont{Romalis, Griffith,
  and Fortson}}]{romalis01}
\bibinfo{author}{\bibfnamefont{M.}~\bibnamefont{Romalis}},
  \bibinfo{author}{\bibfnamefont{W.}~\bibnamefont{Griffith}}, \bibnamefont{and}
  \bibinfo{author}{\bibfnamefont{E.}~\bibnamefont{Fortson}},
  \bibinfo{journal}{Phys.\ Rev.\ Lett.} \textbf{\bibinfo{volume}{86}},
  \bibinfo{pages}{2505} (\bibinfo{year}{2001}).

\bibitem[{\citenamefont{Peccei and Quinn}(1977)}]{peccei77}
\bibinfo{author}{\bibfnamefont{R.}~\bibnamefont{Peccei}} \bibnamefont{and}
  \bibinfo{author}{\bibfnamefont{H.}~\bibnamefont{Quinn}},
  \bibinfo{journal}{Phys.\ Rev.\ Lett.} \textbf{\bibinfo{volume}{38}},
  \bibinfo{pages}{1440} (\bibinfo{year}{1977}).

\bibitem[{\citenamefont{Collar et~al.}(2003)}]{collar03}
\bibinfo{author}{\bibfnamefont{J.}~\bibnamefont{Collar}} \bibnamefont{et~al.},
  \bibinfo{journal}{hep-ex/0304024}  (\bibinfo{year}{2003}),
  \bibinfo{note}{presented at the SPIE conference on Astronomical Telescopes
  and Instrumentation, Waikoloa, Hawaii, 2002}.

\bibitem[{\citenamefont{Tada et~al.}(2001)}]{tada01}
\bibinfo{author}{\bibfnamefont{M.}~\bibnamefont{Tada}} \bibnamefont{et~al.},
  \bibinfo{journal}{physics/0101028}  (\bibinfo{year}{2001}).

\bibitem[{\citenamefont{Abel et~al.}(2001)\citenamefont{Abel, Khalil, and
  Lebedev}}]{abel01}
\bibinfo{author}{\bibfnamefont{S.}~\bibnamefont{Abel}},
  \bibinfo{author}{\bibfnamefont{S.}~\bibnamefont{Khalil}}, \bibnamefont{and}
  \bibinfo{author}{\bibfnamefont{O.}~\bibnamefont{Lebedev}},
  \bibinfo{journal}{Nucl.\ Phys.\ B} \textbf{\bibinfo{volume}{606}},
  \bibinfo{pages}{151} (\bibinfo{year}{2001}).

\bibitem[{\citenamefont{Falk et~al.}(1999)\citenamefont{Falk, Olive, Pospelov,
  and Roiban}}]{falk99}
\bibinfo{author}{\bibfnamefont{T.}~\bibnamefont{Falk}},
  \bibinfo{author}{\bibfnamefont{K.}~\bibnamefont{Olive}},
  \bibinfo{author}{\bibfnamefont{M.}~\bibnamefont{Pospelov}}, \bibnamefont{and}
  \bibinfo{author}{\bibfnamefont{R.}~\bibnamefont{Roiban}},
  \bibinfo{journal}{Nucl.\ Phys.\ B} \textbf{\bibinfo{volume}{560}},
  \bibinfo{pages}{3} (\bibinfo{year}{1999}).

\bibitem[{\citenamefont{Regan et~al.}(2002)\citenamefont{Regan, Commins, Smidt,
  and DeMille}}]{regan02}
\bibinfo{author}{\bibfnamefont{B.}~\bibnamefont{Regan}},
  \bibinfo{author}{\bibfnamefont{E.}~\bibnamefont{Commins}},
  \bibinfo{author}{\bibfnamefont{C.}~\bibnamefont{Smidt}}, \bibnamefont{and}
  \bibinfo{author}{\bibfnamefont{D.}~\bibnamefont{DeMille}},
  \bibinfo{journal}{Phys.\ Rev.\ Lett.} \textbf{\bibinfo{volume}{88}},
  \bibinfo{pages}{071805} (\bibinfo{year}{2002}).

\bibitem[{\citenamefont{Hudson et~al.}(2002)\citenamefont{Hudson, Sauer,
  Tarbutt, and Hinds}}]{hudson02}
\bibinfo{author}{\bibfnamefont{J.}~\bibnamefont{Hudson}},
  \bibinfo{author}{\bibfnamefont{B.}~\bibnamefont{Sauer}},
  \bibinfo{author}{\bibfnamefont{M.}~\bibnamefont{Tarbutt}}, \bibnamefont{and}
  \bibinfo{author}{\bibfnamefont{E.}~\bibnamefont{Hinds}},
  \bibinfo{journal}{Phys.\ Rev.\ Lett.} \textbf{\bibinfo{volume}{89}},
  \bibinfo{pages}{272701} (\bibinfo{year}{2002}).

\bibitem[{\citenamefont{Kozlov and DeMille}(2002)}]{kozlov02}
\bibinfo{author}{\bibfnamefont{M.}~\bibnamefont{Kozlov}} \bibnamefont{and}
  \bibinfo{author}{\bibfnamefont{D.}~\bibnamefont{DeMille}},
  \bibinfo{journal}{Phys.\ Rev.\ Lett.} \textbf{\bibinfo{volume}{89}},
  \bibinfo{pages}{133001} (\bibinfo{year}{2002}).

\bibitem[{\citenamefont{Lebedev et~al.}(2004)\citenamefont{Lebedev, Olive,
  Pospelov, and Ritz}}]{lebedev04}
\bibinfo{author}{\bibfnamefont{O.}~\bibnamefont{Lebedev}},
  \bibinfo{author}{\bibfnamefont{K.}~\bibnamefont{Olive}},
  \bibinfo{author}{\bibfnamefont{M.}~\bibnamefont{Pospelov}}, \bibnamefont{and}
  \bibinfo{author}{\bibfnamefont{A.}~\bibnamefont{Ritz}},
  \bibinfo{journal}{Phys.\ Rev.\ D} \textbf{\bibinfo{volume}{70}},
  \bibinfo{pages}{016003} (\bibinfo{year}{2004}).

\bibitem[{\citenamefont{Abel and Lebedev}(2006)}]{abel06}
\bibinfo{author}{\bibfnamefont{S.}~\bibnamefont{Abel}} \bibnamefont{and}
  \bibinfo{author}{\bibfnamefont{O.}~\bibnamefont{Lebedev}},
  \bibinfo{journal}{JHEP} \textbf{\bibinfo{volume}{0601}}, \bibinfo{pages}{133}
  (\bibinfo{year}{2006}).

\bibitem[{\citenamefont{Ramsey}(1994)}]{ramsey95}
\bibinfo{author}{\bibfnamefont{N.}~\bibnamefont{Ramsey}}, in
  \emph{\bibinfo{booktitle}{Proc.\ XIV Int.\ Conf.\ Atomic Physics}}
  (\bibinfo{publisher}{AIP}, \bibinfo{address}{New York},
  \bibinfo{year}{1994}), p.~\bibinfo{pages}{3}.

\bibitem[{\citenamefont{Barr}(1993)}]{barr93a}
\bibinfo{author}{\bibfnamefont{S.}~\bibnamefont{Barr}}, \bibinfo{journal}{Int.\
  J.\ Mod.\ Phys.\ A} \textbf{\bibinfo{volume}{8}}, \bibinfo{pages}{209}
  (\bibinfo{year}{1993}).

\bibitem[{\citenamefont{Khriplovich and
  Lamoreaux}(1996)}]{khriplovich_lamoreaux}
\bibinfo{author}{\bibfnamefont{I.}~\bibnamefont{Khriplovich}} \bibnamefont{and}
  \bibinfo{author}{\bibfnamefont{S.}~\bibnamefont{Lamoreaux}},
  \emph{\bibinfo{title}{CP Violation without Strangeness}}
  (\bibinfo{publisher}{Springer}, \bibinfo{year}{1996}).

\bibitem[{ibr()}]{ibrahim07}
\bibinfo{note}{ArXiv:0705.2008 [hep-ph]}.

\bibitem[{\citenamefont{Ramsey}(1956)}]{ramsey_molec_beams}
\bibinfo{author}{\bibfnamefont{N.}~\bibnamefont{Ramsey}},
  \emph{\bibinfo{title}{Molecular Beams}} (\bibinfo{publisher}{Oxford
  University Press}, \bibinfo{year}{1956}).

\bibitem[{\citenamefont{Zel$^\prime$dovich}(1959)}]{zeldovich59}
\bibinfo{author}{\bibfnamefont{Y.}~\bibnamefont{Zel$^\prime$dovich}},
  \bibinfo{journal}{Sov.\ Phys.\ JETP} \textbf{\bibinfo{volume}{9}},
  \bibinfo{pages}{1389} (\bibinfo{year}{1959}).

\bibitem[{\citenamefont{Fermi and Zinn}(1946)}]{fermi46}
\bibinfo{author}{\bibfnamefont{E.}~\bibnamefont{Fermi}} \bibnamefont{and}
  \bibinfo{author}{\bibfnamefont{W.}~\bibnamefont{Zinn}},
  \bibinfo{journal}{Phys.\ Rev.} \textbf{\bibinfo{volume}{70}},
  \bibinfo{pages}{103} (\bibinfo{year}{1946}).

\bibitem[{\citenamefont{Byrne}(1993)}]{byrne}
\bibinfo{author}{\bibfnamefont{J.}~\bibnamefont{Byrne}},
  \emph{\bibinfo{title}{Neutrons, Nuclei and Matter}}
  (\bibinfo{publisher}{Institute of Physics Publishing}, \bibinfo{year}{1993}).

\bibitem[{\citenamefont{Golub et~al.}(1991)\citenamefont{Golub, Richardson, and
  Lamoreaux}}]{golub_UCN_book}
\bibinfo{author}{\bibfnamefont{R.}~\bibnamefont{Golub}},
  \bibinfo{author}{\bibfnamefont{D.}~\bibnamefont{Richardson}},
  \bibnamefont{and}
  \bibinfo{author}{\bibfnamefont{S.}~\bibnamefont{Lamoreaux}},
  \emph{\bibinfo{title}{Ultra-Cold Neutrons}} (\bibinfo{publisher}{Adam
  Hilger}, \bibinfo{address}{Bristol}, \bibinfo{year}{1991}).

\bibitem[{\citenamefont{Altarev et~al.}(1996)\citenamefont{Altarev, Borisov,
  Borikova et~al.}}]{altarev96}
\bibinfo{author}{\bibfnamefont{I.}~\bibnamefont{Altarev}},
  \bibinfo{author}{\bibfnamefont{Y.}~\bibnamefont{Borisov}},
  \bibinfo{author}{\bibfnamefont{N.}~\bibnamefont{Borikova}},
  \bibnamefont{et~al.}, \bibinfo{journal}{Phys.\ Atomic Nuclei}
  \textbf{\bibinfo{volume}{59}}, \bibinfo{pages}{1152} (\bibinfo{year}{1996}).

\bibitem[{\citenamefont{Fedorov et~al.}(2001)\citenamefont{Fedorov, Lapin,
  Semenikhin, and Voronin}}]{fedorov01}
\bibinfo{author}{\bibfnamefont{V.}~\bibnamefont{Fedorov}},
  \bibinfo{author}{\bibfnamefont{E.}~\bibnamefont{Lapin}},
  \bibinfo{author}{\bibfnamefont{S.}~\bibnamefont{Semenikhin}},
  \bibnamefont{and} \bibinfo{author}{\bibfnamefont{V.}~\bibnamefont{Voronin}},
  \bibinfo{journal}{Physica B: Physics of Condensed Matter}
  \textbf{\bibinfo{volume}{297}}, \bibinfo{pages}{293} (\bibinfo{year}{2001}).

\bibitem[{\citenamefont{Steyerl et~al.}(1986)\citenamefont{Steyerl, Nagel,
  Schreiber et~al.}}]{steyerl86}
\bibinfo{author}{\bibfnamefont{A.}~\bibnamefont{Steyerl}},
  \bibinfo{author}{\bibfnamefont{H.}~\bibnamefont{Nagel}},
  \bibinfo{author}{\bibfnamefont{F.-X.} \bibnamefont{Schreiber}},
  \bibnamefont{et~al.}, \bibinfo{journal}{Phys.\ Lett.\ A}
  \textbf{\bibinfo{volume}{116}}, \bibinfo{pages}{347} (\bibinfo{year}{1986}).

\bibitem[{\citenamefont{Golub and Pendlebury}(1977)}]{golub_pendlebury77}
\bibinfo{author}{\bibfnamefont{R.}~\bibnamefont{Golub}} \bibnamefont{and}
  \bibinfo{author}{\bibfnamefont{J.}~\bibnamefont{Pendlebury}},
  \bibinfo{journal}{Phys.\ Lett. A} \textbf{\bibinfo{volume}{62}},
  \bibinfo{pages}{337} (\bibinfo{year}{1977}).

\bibitem[{\citenamefont{Landau}(1941)}]{landau41}
\bibinfo{author}{\bibfnamefont{L.}~\bibnamefont{Landau}}, \bibinfo{journal}{J.\
  Phys.\ U.S.S.R.} \textbf{\bibinfo{volume}{5}}, \bibinfo{pages}{71}
  (\bibinfo{year}{1941}).

\bibitem[{\citenamefont{Landau}(1947)}]{landau47}
\bibinfo{author}{\bibfnamefont{L.}~\bibnamefont{Landau}}, \bibinfo{journal}{J.\
  Phys.\ U.S.S.R.} \textbf{\bibinfo{volume}{11}}, \bibinfo{pages}{91}
  (\bibinfo{year}{1947}).

\bibitem[{\citenamefont{Feynman}(1954)}]{feynman54}
\bibinfo{author}{\bibfnamefont{R.}~\bibnamefont{Feynman}},
  \bibinfo{journal}{Phys.\ Rev.} \textbf{\bibinfo{volume}{91}},
  \bibinfo{pages}{1291} (\bibinfo{year}{1954}).

\bibitem[{\citenamefont{Yarnell et~al.}(1959)\citenamefont{Yarnell, Arnold,
  Bendt, and Kerr}}]{yarnell59}
\bibinfo{author}{\bibfnamefont{J.}~\bibnamefont{Yarnell}},
  \bibinfo{author}{\bibfnamefont{G.}~\bibnamefont{Arnold}},
  \bibinfo{author}{\bibfnamefont{P.}~\bibnamefont{Bendt}}, \bibnamefont{and}
  \bibinfo{author}{\bibfnamefont{E.}~\bibnamefont{Kerr}},
  \bibinfo{journal}{Phys.\ Rev.} \textbf{\bibinfo{volume}{113}},
  \bibinfo{pages}{1379} (\bibinfo{year}{1959}).

\bibitem[{\citenamefont{Cohen and Feynman}(1957)}]{cohen57}
\bibinfo{author}{\bibfnamefont{M.}~\bibnamefont{Cohen}} \bibnamefont{and}
  \bibinfo{author}{\bibfnamefont{R.}~\bibnamefont{Feynman}},
  \bibinfo{journal}{Phys.\ Rev.} \textbf{\bibinfo{volume}{107}},
  \bibinfo{pages}{13} (\bibinfo{year}{1957}).

\bibitem[{sns()}]{sns_edm}
\bibinfo{note}{http://p25ext.lanl.gov/edm/edm.html}.

\bibitem[{psi({\natexlab{a}})}]{psi_ucn}
\bibinfo{note}{http://ucn.web.psi.ch/index.htm}.

\bibitem[{psi({\natexlab{b}})}]{psi_edm}
\bibinfo{note}{http://nedm.web.psi.ch/}.

\bibitem[{ram()}]{ramsey_private_michelson}
\bibinfo{note}{N.F. Ramsey, private communication}.

\bibitem[{\citenamefont{Green et~al.}(1998)\citenamefont{Green, Harris,
  Iaydjiev et~al.}}]{green98}
\bibinfo{author}{\bibfnamefont{K.}~\bibnamefont{Green}},
  \bibinfo{author}{\bibfnamefont{P.}~\bibnamefont{Harris}},
  \bibinfo{author}{\bibfnamefont{P.}~\bibnamefont{Iaydjiev}},
  \bibnamefont{et~al.}, \bibinfo{journal}{Nucl.\ Instr.\ Meth. A}
  \textbf{\bibinfo{volume}{404}}, \bibinfo{pages}{381} (\bibinfo{year}{1998}).

\bibitem[{\citenamefont{Bloch and Siegert}(1940)}]{bloch40}
\bibinfo{author}{\bibfnamefont{F.}~\bibnamefont{Bloch}} \bibnamefont{and}
  \bibinfo{author}{\bibfnamefont{A.}~\bibnamefont{Siegert}},
  \bibinfo{journal}{Phys.\ Rev.} \textbf{\bibinfo{volume}{57}},
  \bibinfo{pages}{522} (\bibinfo{year}{1940}).

\bibitem[{\citenamefont{Ramsey}(1955)}]{ramsey55}
\bibinfo{author}{\bibfnamefont{N.}~\bibnamefont{Ramsey}},
  \bibinfo{journal}{Phys.\ Rev.} \textbf{\bibinfo{volume}{100}},
  \bibinfo{pages}{1191} (\bibinfo{year}{1955}).

\bibitem[{\citenamefont{Pendlebury et~al.}(2004)}]{pendlebury04}
\bibinfo{author}{\bibfnamefont{J.}~\bibnamefont{Pendlebury}}
  \bibnamefont{et~al.}, \bibinfo{journal}{Phys.\ Rev.\ A}
  \textbf{\bibinfo{volume}{70}}, \bibinfo{pages}{032102}
  (\bibinfo{year}{2004}).

\bibitem[{\citenamefont{Lamoreaux and Golub}(2007)}]{lamoreaux07}
\bibinfo{author}{\bibfnamefont{S.}~\bibnamefont{Lamoreaux}} \bibnamefont{and}
  \bibinfo{author}{\bibfnamefont{R.}~\bibnamefont{Golub}},
  \bibinfo{journal}{Phys.\ Rev.\ Lett.} \textbf{\bibinfo{volume}{98}},
  \bibinfo{pages}{149101} (\bibinfo{year}{2007}).

\bibitem[{\citenamefont{Harris and Pendlebury}(2006)}]{harris06}
\bibinfo{author}{\bibfnamefont{P.}~\bibnamefont{Harris}} \bibnamefont{and}
  \bibinfo{author}{\bibfnamefont{J.}~\bibnamefont{Pendlebury}},
  \bibinfo{journal}{Phys. Rev. A} \textbf{\bibinfo{volume}{73}},
  \bibinfo{pages}{014101} (\bibinfo{year}{2006}).

\bibitem[{\citenamefont{Yoshiki et~al.}(1992)}]{yoshiki92}
\bibinfo{author}{\bibfnamefont{H.}~\bibnamefont{Yoshiki}} \bibnamefont{et~al.},
  \bibinfo{journal}{Phys.\ Rev.\ Lett.} \textbf{\bibinfo{volume}{68}},
  \bibinfo{pages}{1323} (\bibinfo{year}{1992}).

\bibitem[{\citenamefont{Baker et~al.}(2003{\natexlab{a}})}]{baker03a}
\bibinfo{author}{\bibfnamefont{C.}~\bibnamefont{Baker}} \bibnamefont{et~al.},
  \bibinfo{journal}{Nucl.\ Instr.\ Meth.\ A} \textbf{\bibinfo{volume}{501}},
  \bibinfo{pages}{517} (\bibinfo{year}{2003}{\natexlab{a}}).

\bibitem[{\citenamefont{Baker et~al.}(2003{\natexlab{b}})}]{baker03b}
\bibinfo{author}{\bibfnamefont{C.}~\bibnamefont{Baker}} \bibnamefont{et~al.},
  \bibinfo{journal}{Physics Letters A} \textbf{\bibinfo{volume}{308-1}},
  \bibinfo{pages}{67} (\bibinfo{year}{2003}{\natexlab{b}}).

\bibitem[{\citenamefont{Yoshiki}(1996)}]{yoshiki96}
\bibinfo{author}{\bibfnamefont{H.}~\bibnamefont{Yoshiki}}, in
  \emph{\bibinfo{booktitle}{Proceedings of the International Workshop on Future
  Prospects of Baryon Instability Search in p-decay and n-$\bar{\rm n}$
  Oscillation Experiments}}, edited by
  \bibinfo{editor}{\bibfnamefont{S.}~\bibnamefont{Ball}} \bibnamefont{and}
  \bibinfo{editor}{\bibfnamefont{Y.}~\bibnamefont{Kamyshkov}}
  (\bibinfo{publisher}{Oak Ridge, Tennessee, U.S.A.}, \bibinfo{year}{1996}), p.
  \bibinfo{pages}{345}.

\bibitem[{cry()}]{cryoedm_proposal}
\bibinfo{note}{http://arxiv.org/abs/0709.2428}.

\bibitem[{\citenamefont{Serebrov et~al.}(2005)}]{serebrov_edm}
\bibinfo{author}{\bibfnamefont{A.}~\bibnamefont{Serebrov}}
  \bibnamefont{et~al.}, \bibinfo{journal}{J.\ Res.\ Natl.\ Inst.\ Stand.\
  Technol.} \textbf{\bibinfo{volume}{110}}, \bibinfo{pages}{185}
  (\bibinfo{year}{2005}).

\bibitem[{coo()}]{cooper}
\bibinfo{note}{M. Cooper et al.; http://p25ext.lanl.gov/edm/edm.html}.

\bibitem[{\citenamefont{Golub and Lamoreaux}(1994)}]{golub94}
\bibinfo{author}{\bibfnamefont{R.}~\bibnamefont{Golub}} \bibnamefont{and}
  \bibinfo{author}{\bibfnamefont{S.}~\bibnamefont{Lamoreaux}},
  \bibinfo{journal}{Phys.\ Rep.} \textbf{\bibinfo{volume}{237}},
  \bibinfo{pages}{1} (\bibinfo{year}{1994}).


\end{thebibliography}

\end{document}